\documentclass[fleqn,twoside]{article}%
\usepackage{amsmath}
\usepackage{amsfonts}
\usepackage{amssymb}
\usepackage{graphicx}
\usepackage{cite}
\def\be{\begin{equation}}

\def\ee{\end{equation}}
\def\bes{\begin{equation}\begin{split}&}
\def\es{\end{split}}
\def\bi{\bibitem}

\begin{document}
\title{Degenerate Hamiltonian operator in higher-order canonical gravity - the problem and a remedy}
\author{Abhik Kumar Sanyal}

\maketitle
\noindent
\begin{center}
\noindent
Dept. of Physics, Jangipur College, Murshidabad,
\noindent
West Bengal, India - 742213. \\

\end{center}
\footnotetext[1] {\noindent
Electronic address:\\
\noindent
sanyal\_ ak@yahoo.com}
\noindent
\abstract{Different routes towards canonical formulation of a classical theory result in different canonically equivalent Hamiltonians, while their quantum counterparts are related through appropriate unitary transformation. However, for higher-order theory of gravity although two Hamiltonians emerging from the same action differing by total derivative terms are related through canonical transformation, the difference transpires while attempting canonical quantization, which is predominant in non-minimally coupled higher-order theory of gravity. We follow Dirac's constraint analysis to formulate phase-space structures, in the presence (case-I) and absence (case-II) of total derivative terms. While the coupling parameter plays no significant role as such for case-I, quantization depends on its form explicitly in case-II, and as a result unitary transformation relating the two is not unique. We find certain mathematical inconsistency in case-I, for modified Gauss-Bonnet-Dilatonic coupled action, in particular. Thus, we conclude that total derivative terms indeed play a major role in the quantum domain and should be taken care of a-priori, for consistency.}\\

\noindent
keywords: Higher Order theory; Canonical Formulation; Total derivative terms; Canonical quantization.

\maketitle

\flushbottom

\section{Introduction}

One cannot avoid the presence of higher order curvature invariant terms when gravity is strong enough, particularly in the very early universe or in the vicinity of a black-hole. Naturally, canonical formulation of such theories is a very important issue to study, which is of-course non-trivial. Since, a fourth order field equation may be cast into two second order ones in view of an additional degree of freedom, so canonical formulation of higher order theory of gravity with curvature squared term may be performed in view of the basic variables, viz., the three-space metric $h_{ij}$ and the extrinsic curvature tensor $K_{ij}$. The oldest technique in this regard was developed long ago by Ostrogradski \cite{1O, 1O1}. However, if the Hessian determinant vanishes and the Lagrangian becomes singular, e.g. in the presence of lapse function ($N$), Ostrogradski's technique does not work and it is required to follow Dirac's algorithm of constrained analysis \cite{2aD, 2bD}. It is possible to bypasses the constrained analysis following Horowitz formalism (HF) \cite{3H}. In HF, instead of extrinsic curvature tensor, it is required to introduce auxiliary variable at the beginning (which is the first derivative of the action with respect to the highest derivative of the field variable that appears in the action) judiciously into the action, so that it is canonical. Phase-space structure is finally obtained upon translation to the basic variable ($K_{ij}$) through canonical transformation. In this connection let us mention that under variational technique, total derivative terms are removed from the action. Whether these terms vanish at the boundary (end points) or supplementary boundary terms are required to mutually cancel them, depends on the choice of boundary data. Once Ostrogradski's technique or Dirac's algorithm is initiated by a change of variable, there is absolutely no further scope to remove the total derivative terms from the action. On the contrary, one can integrate the action by parts before initiating these formalisms. Eventually, two different Hamiltonian results, which are of-course canonically equivalent. In Horowitz' formalism (HF) on the other hand, the action is first expressed in terms of the auxiliary variable and the total derivative terms are then integrated out by parts. All the techniques mentioned so far, which involve integration by parts at some stage, tacitly assume $\delta{h_{ij}}|_{\partial\mathcal{V}} =  0 = \delta{K_{ij}}|_{\partial\mathcal{V}}$ at the boundary. As a result, the total derivative terms vanish at the boundary, and the actions are devoid of supplementary boundary terms. It is important to note that since the variation of Gibbons-Hawking-York term \cite{4aGHY, 4bGHY} also vanishes, it is no longer mandatary to retain such term. \\

A few problems with HF were noticed sometimes back. Particularly, it was shown that HF allows to introduce auxiliary variable even in linear theory of gravity, which results in wrong quantum equation, being different from the standard Wheeler-de-Witt equation \cite{P, 5aMH, 5bMH, 5cMH, 5dMH, 5gMH}. In view of this, yet another treatment towards canonical formulation of higher order theory of gravity had therefore been developed by the name ``Modified Horowitz' Formalism" (MHF) \cite{5aMH, 5bMH, 5cMH, 5dMH, 5gMH, 5eMH, 5fMH,  br2, 5jMH, 5kMH}. The primary aspect of MHF is to integrate the action by parts prior to the installation of auxiliary variable. Now if one starts from an action being expressed in terms of the scale factor, as proposed by Horowitz \cite{3H}, there is a possibility of eliminating some additional total derivative terms which do not appear from the variational principle \cite{5dMH, 5fMH}. To avoid such uncanny situation, MHF proposes to express the action in terms of the basic variable, viz. the three space metric $h_{ij}$. Further, in MHF it is customary to set $\delta{h_{ij}}|_{\partial\mathcal{V}} =  0 = \delta{R}|_{\partial\mathcal{V}}$ at the boundary, $R$ being the $4$-dimensional Ricci scalar. The reason being, under conformal transformation or following some appropriate redefinition, $F(R)$ theory may be translated to equivalent scalar-tensor theories in Einstein's or Jordan's frame respectively. Euler-Lagrange equations may then be derived provided the scalar field vanishes at the end points. This is equivalent to the vanishing of the $4$ - dimensional Ricci scalar ($R$) at the boundary. However, under such choice, total derivative terms do not vanish at the end points, and therefore the action is supplemented by appropriate boundary terms. In the process, it is therefore also possible to set higher-order curvature invariant term vanish at any stage, resulting in General theory of Relativity associated with Gibbons-Hawking-York term. So the scheme in MHF is to express the action in terms of $h_{ij}$, and to split the higher-order supplementary boundary terms into two parts. The action is integrated by parts, and some of the total derivative terms are mutually cancelled with supplementary boundary terms. Next, auxiliary variable, following the suggestion of Horowitz is installed, and the action is integrated by parts yet again, to mutually cancel the total derivative terms with the rest of the supplementary boundary terms. It is to be mentioned that in MHF, the choice $\delta{h_{ij}}|_{\partial\mathcal{V}} =  0 = \delta{K_{ij}}|_{\partial\mathcal{V}}$ or $\delta{h_{ij}}|_{\partial\mathcal{V}} =  0 = \delta{R}|_{\partial\mathcal{V}}$ is just a matter of taste, since it does not affect the phase-space structure. In fact, since $R$ is not treated as a variable of the theory, so, instead of $R$, one can even fix the auxiliary variable at the boundary too.\\

Now, it is generally believed that total derivative term appearing in an action, does not affect the equation of motion, rather it just changes the canonical momenta in a way that only amounts to a canonical transformation.  Main purpose of the present manuscript is to explore the effect of total derivative terms on the phase-space structures of non-minimally coupled higher-order theories of gravity.
Even if the action is expressed in terms of $h_{ij}$ a-priori, and the same boundary choice are invoked, the two formalisms (HF and MHF) differ. As mentioned, the difference lies in the fact that in HF, the auxiliary variable is installed and thereafter the action is integrated by parts, while in MHF, it is done otherwise. Such difference transpires from the two Hamiltonian obtained following the two routes even in minimally coupled higher-order theory. Nevertheless, the two are related under canonical transformation \cite{5jMH}, and therefore there is absolutely no problem in the classical domain. In fact, if Dirac formalism is followed right from an action, without integrating it by parts (Case-I), it leads to the phase-space structure as in the case of HF. On the contrary, if the same is done after integrating the action by parts and thus removing total derivative terms from the action (Case-II), the resulting Hamiltonian is equivalent to the one obtainable following MHF. However, in the present manuscript we show that the two canonically equivalent Hamiltonians yield completely different quantum descriptions in the case of non-minimal higher-order theories, leading to the so-called `Degenerate Hamiltonian operator'. It is already known that the same classical Hamiltonian equations can lead to the different quantum theories, depending on the type of the symbol. The obstructions to the equivalence of different quantizations are possible, while they are connected to the topology of the phase space. If it is $T^* \mathbb{R}^n$, all the quantizations are unitary equivalent of the same classical theory. It is just the change of the variables in the wave function and the phase transformation, plus the change of the integration measure, and the transformation of the momenta respecting the change of the measure. It is the unitary transformation. Else, there exists quantum canonical transformations for non-unitary transformations \cite{Arlen}. Nevertheless, for non-minimal higher-order theories, canonical quantization is not affected by the coupling parameter for case-I, while, different forms yield different quantum descriptions for case-II. Thus, such transformations are not one to one in the case of non-minimal theories. Further, case-I eliminates additional boundary terms from the action, which is not supported by variational principle. Such mathematical inconsistency appears in the modified Gauss-Bonnet-Dilatonic coupled action (MEGBD) in Case-I, when HF equivalent to case-I is invoked. This fact direct us to pick up Case-II as the correct description required for quantization, and in the process such degeneracy is removed.\\

In the following section, we take up three systems described by non-minimally coupled higher-order theories, which are special cases of $f(\phi, R)$, and $f(\phi, R, \mathcal{G})$ theories ($\mathcal{G}$ being the Gauss-Bonnet term), and perform canonical formulation in the background of Robertson-Walker metric following Dirac's analysis, first keeping the total derivative terms intact (Case-I), and second removing the same from the action (Case-II). The viability of both the canonically equivalent Hamiltonians are then tested in the classical and quantum domains. We perform detailed mathematical computation, which might appear to be the run of the machine. Nevertheless, it is performed to establish mathematical rigor. While no irregularities appear in the first two cases, the Hamiltonian for MEGBD in Case-I, shows certain inconsistency. In section 3, we follow HF and MHF to explore the fact that Case-I is identical to HF, while Case-II leads to MHF. Additional inconsistency appears in the case of MEGBD in HF. We conclude in section 4.

\section{Canonical formulation of non-minimally coupled higher-order theory of gravity}

In a recent article \cite{5jMH}, it has been shown that for the minimally coupled higher-order gravitational action given in the form,
$A = \int[\alpha R + \beta R^2 - {1\over 2}\phi_{;\mu}\phi^{;\mu} - V(\phi)]\sqrt{-g} d^4 x$, the two Hamiltonians obtained following HF and MHF, are related under canonical transformation. However, the two Hamiltonians lead to two different quantum descriptions. Particularly, while extremizing the effective potential obtained under quantization of the Hamiltonian followed from MHF leads to inflation, no such result is revealed in the HF. Further, although the on-shell semiclassical wave-functions show oscillatory behaviour about the classical inflationary solution in both the case, neither the exponents nor the pre-factors is the same. Such difference is predominant in the case of non-minimal coupling, which we explore here. The whole analysis is performed in the Robertson-Walker mini-superspace background,

\be \label{RW} ds^2 = -N(t)^2 dt^2 + a(t)^2 \left[ {dr^2\over 1-kr^2} + r^2 d\theta^2 - r^2 \sin^2{\theta} d\phi^2\right],\ee
in which the basic variables are the induced three-space metric, $h_{ij} =  a^2 \delta_{ij} = z\delta_{ij}$, and the extrinsic curvature tensor, $K_{ij} = -{\dot h_{ij}\over 2 N } =- {a\dot a\over N}\delta_{ij} = -{\dot z\over N} \delta_{ij} = - x\delta_{ij}$, so that

\be\label{x}  x = {\dot z\over N}.\ee
The Ricci scalar in connection with the metric \eqref{RW} is expressed in terms of the basic variable $z= a^2$, as

\be \label{R1} R = \frac{6}{N^2}\left(\frac{\ddot a}{a}+\frac{\dot a^2}{a^2}+N^2\frac{k}{a^2}-\frac{\dot N\dot a}{N a}\right)
 = {6\over N^2}\left[{\ddot z\over 2z} + N^2 {k\over z} - {1\over 2}{\dot N\dot z\over N z}\right].\ee
In the following three subsections, we study non-minimally coupled scalar-tensor theory of gravity in the presence of scalar
curvature squared term, scalar-tensor theory of gravity in the presence of non-minimally coupled higher-order term, and higher-order theory of gravity in the presence of Gauss-Bonnet term with Dilatonic coupling respectively, following Dirac's algorithm for constrained system. It is important to mention that, if the action is expressed in terms of the basic variables $h_{ij}$ and $K_{ij}$ from the very beginning, the theory reduces to lower order with additional degrees of freedom, and neither HF nor MHF may be applied. Ostrogradski's technique works provided the Hessian determinant doesn't vanish. Otherwise, as in the present situations, one has to perform constraint analysis, following Dirac's algorithm. If an action represents as higher order theory of gravity, one can immediately express it in terms of $h_{ij}$ and $K_{ij}$ and initiate Dirac's algorithm (Case-I). On the other hand, one can first integrate the action by parts to remove the total derivative terms and then express the action in terms of the basic variables and initiate Dirac's formalism (Case-II) thereafter. In the following subsections we construct the phase-space structures in connection with all the three actions following Case-I and Case-II, explore canonical equivalence, and perform canonical quantization. In the process, we expatiating the fact that although the two Hamiltonians in each case lead to viable quantum dynamics, they are different altogether. This means, there is no unique quantum description of a classical theory. This is what we mean by the degeneracy of the Hamiltonian operator. From the view point of mathematical consistency, we also suggest a remedy.

\subsection{Scalar-tensor theory of gravity in the presence of higher-order term}

In a recent article \cite{5kMH}, the following $f(\phi, R)$ gravitational action

\be\label{Anm1} A_1 = \int\left[f(\phi) R + \beta R^2 - {1\over 2}\phi_{,\mu}\phi^{,\mu} - V(\phi)\right]\sqrt{-g}~d^4x.
\ee
has been extensively studied in the context of the evolution of the early universe. In the minisuperspace \eqref{RW} under consideration, the above action \eqref{Anm1} is expressed as

\be\begin{split}\label{Anm2}
A_1 &= \int\bigg[{3 f\sqrt z}\Big(\frac{\ddot z}{ N}- \frac{\dot N \dot z}{N^2} + 2k N \Big)+\frac{9 \beta}{\sqrt z}
\Big(\frac{{\ddot z}^2}{N^3} -
\frac{2 \dot N \dot z \ddot z}{N^4} + \frac{{\dot N}^2{\dot z}^2}{N^5} -\frac{4k\dot N \dot z}{N^2}
+ \frac{4 k {\ddot z}}{N} + 4 k^2 N \Big)\\&\hspace{2.0 in}+z^{\frac{3}{2}} \Big(\frac{1}{2N}\dot\phi^2-VN\Big)\bigg]dt.
\end{split}\ee
Field equations are found following variational principle, and the total derivative terms, $\Sigma_R =f(\phi)\frac{3\sqrt z\dot z}{N},~~\text{and}~~\Sigma_{R^2} = \Sigma_{R^2_1}+ \Sigma_{R^2_2}, ~~\text{where}~~\Sigma_{R^2_1}= \frac{36 \beta k\dot z}{N\sqrt z}~~\text{and}  ~~\Sigma_{R^2_2} =\frac{18 \beta \dot z}{N^3\sqrt z}\left({\ddot z}-\frac{\dot z\dot N}{N} \right)$ vanish at the boundary, due to the choice $\delta h_{ij}|_{\mathcal{\partial V}} = 0 = \delta K_{ij}|_{\mathcal{\partial V}}$ \footnote{One can instead add these terms with revered sign in the action, which we call supplementary boundary terms, and choose $\delta h_{ij}|_{\mathcal{\partial V}} = 0 = \delta R|_{\mathcal{\partial V}}$ at the boundary, as mentioned in the introduction}. The scalar field equation and the $(^0_0)$ component of Einstein's equation are,

\be\begin{split}\label{00}
~& \ddot\phi + \left(3{\dot a\over a} - {\dot N\over N}\right)\dot\phi + {V'\over N^2} - 6f'\left({\ddot a\over a}
+ {\dot a^2\over a^2} - {\dot N\over N}{\dot a\over a} + k{N^2\over a^2}\right) = 0,\\&\Bigg[{6f\over a^2}
\left({\dot a^2\over N^2}+k\right)+6{f'\dot a\dot\phi\over N^2 a} +{36\beta\over a^2N^4}
\Bigg(2\dot a\dddot a -\ddot a^2 +2{\dot a^2\ddot a\over a} - 3{\dot a^4\over a^2}
- 2\dot a^2{\ddot N\over N} - 4{\dot N\over N} \dot a\ddot a \\&\hspace{1.2 in}+ 5 \dot a^2{\dot N^2\over N^2}
- 2 {\dot a^3 \dot N\over a N} - 2k N^2{\dot a^2\over a^2} +k^2{N^4\over a^2}\Bigg) - \left({\dot \phi^2\over 2 N^2}
+ V(\phi)\right)\Bigg]Na^3=0. \end{split}\ee
$N$ being a Lagrangian multiplier, the above set of field equations reduce to

\be\begin{split}\label{01}
~& \ddot\phi + 3{\dot a\over a}\dot\phi + V' - 6f'\left({\ddot a\over a} + {\dot a^2\over a^2} + {k\over a^2}\right) = 0,
\\&6f\left({\dot a^2\over a^2}+{k\over a^2}\right)+6{f'\dot a\dot\phi\over a} +{36\beta\over a^2}\Bigg(2\dot a\dddot a
-\ddot a^2 +2{\dot a^2\ddot a\over a} - 3{\dot a^4\over a^2}  - 2k {\dot a^2\over a^2} +{k^2\over a^2}\Bigg) -
\left({\dot \phi^2\over 2} + V(\phi)\right)=0. \end{split}\ee
In the above, prime denotes derivative with respect to $\phi$. We do not write the space-space component of Einstein's equation, since it is not an independent one. The above field equations \eqref{01} admit the classical inflationary solution (for $k = 0$) in the form

\be\begin{split} \label{sol1} &a(t) = a_0 e^{\Lambda t},~~~~\phi(t) = \phi_0 e^{-\Lambda t},~~~
\mathrm{under ~the~ condition,}\\&
 V(\phi) = V_1 + {V_0\over \phi},~~f(\phi) = f_0 + {f_1\over \phi} - {\phi^2\over 12},\;\;\;
\mathrm{where},\;\;V_0 = 12 f_1 \Lambda^2,~~V_1 = 6f_0\Lambda^2.\end{split}\ee
In the process, the forms of the coupling parameter $f(\phi)$ as well as the potential $V(\phi)$ have been found,
which we shall require at a later stage.\\

Here we seek phase-space structure of the action \eqref{Anm1}, whose counterpart \eqref{Anm2} in Robertson-Walker metric, is already expressed in terms of $h_{ij} = a^2 \delta_{ij} = z\delta_{ij}$. In the following, we express \eqref{Anm2} in terms of $K_{ij} =  \delta_{ij} = \frac{\dot z}{N}\delta_{ij} = x\delta_{ij}$, and follow Dirac's algorithm. Next, we integrate action \eqref{Anm2} by parts and follow Dirac's algorithm, yet again.

\subsubsection{Hamiltonian, without integrating the action by parts (Case-I):}

To study the phase-space structure of action \eqref{Anm1} following Dirac's algorithm, let us first make change of variable $x = {\dot z\over N}$ in the action \eqref{Anm2}, without controlling the total derivative terms. Treating $({\dot z\over N} - x)$ as a constraint, it is then inserted through a Lagrange multiplier $\lambda$ in the associated point Lagrangian to obtain,

\be\label{DL1} L_1 = 3f\sqrt z(\dot x + 2 k N) + {9\beta\over N\sqrt z} (\dot x + 2k N)^2 + z^{3\over 2}
\left({\dot\phi^2\over 2 N} - NV\right) + \lambda\left({\dot z\over N} - x\right).\ee
Thus, momenta are found as

\be\label{mom1} p_x = 3 f\sqrt z + {18\beta \over N\sqrt z}(\dot x + 2k N); ~~p_\phi = {\dot\phi\over N} z^{3\over 2};
~~p_z = {\lambda\over N};~~p_N = 0 = p_\lambda, \ee
which involve three primary second class constraints. Analysing the constraints appropriately following Dirac's algorithm,
the phase-space structure of the Hamiltonian may be computed as (see appendix B.2 of reference \cite{5kMH})

\be\label{HD1} H_{D1} = N\mathcal{H}_{D1} = N\left[x p_z + {\sqrt z\over 36\beta} p_x^2 -\left({f(\phi) z\over 6\beta}
+ 2k\right)p_x  +  {p_\phi^2\over 2z^{3\over 2}} + {f^2 z^{3\over 2}\over 4\beta} + V z^{3\over 2}\right].\ee
The viability of the above Hamiltonian has been tested in appendix A.\\

\noindent
\textbf{Canonical quantization:}\\

\noindent
Standard canonical quantization of the Hamiltonian \eqref{HD1} leads to,

\be\label{Dq} {i\hbar\over \sqrt z}{\partial \Psi\over \partial z} =  -{\hbar^2\over 36 \beta x}
\left({\partial^2\over \partial x^2} +{n\over x}{\partial\over \partial x}\right)\Psi -
{\hbar^2\over 2 x z^2}\left({\partial^2\Psi\over \partial \phi^2}\right) + {i\hbar\over 2}\left({2k\over \sqrt z}
+{f\sqrt z\over 6 \beta}\right)\left({2\over x}{\partial\Psi\over \partial x}-{\Psi\over x^2}\right)+{z\over x}
\left(V +{f^2\over 4\beta}\right)\Psi.\ee
In the above, Weyl symmetric operator ordering has been performed in the 1st. and the 3rd. terms appearing on right hand side, $n$ being the operator ordering index. Under a further change of variable, the above modified Wheeler-de-Witt equation, takes the look of Schr\"odinger equation, viz.,

\be\label{Ds} i\hbar \frac{\partial \Psi}{\partial\sigma}=-\frac{\hbar^2}{54\beta}
\left(\frac{1}{x}\frac{\partial^2}{\partial x^2} + \frac{n}{x^2}\frac{\partial}{\partial x}\right)
\Psi-\frac{\hbar^2}{3x\sigma^{\frac{4}{3}}}\frac{\partial^2\Psi}{\partial\phi^2}
+i\hbar\left(\frac{4k}{3\sigma^{\frac{1}{3}}}+\frac{f\sigma^{\frac{1}{3}}}
{9\beta}\right)\left(\frac{1}{x}\frac{\partial}{\partial x}-\frac{1}{2 x^2}\right) \Psi+V_e\Psi=\hat{H_{e}}\Psi\ee
where, the proper volume, $\sigma=z^{\frac{3}{2}}=a^3$ plays the role of internal time parameter. In the above,
$\hat H_{e}$ is the effective Hamiltonian operator and $V_e={2\sigma^{\frac{2}{3}}\over 3x}(V+\frac{f^2 }{4\beta})$
is the effective potential. Finally, under appropriate semiclassical approximation, one obtains (see Appendix A)
\be\label{psi1} \Psi_{1} = \Psi_{01} e^{{i\over \hbar}\Lambda\left[\big(48\beta \Lambda^2 + 2 f_0\big)z^{3\over 2}
- {a_0^2\phi_0^2 \sqrt z \over 2}\right]}, ~~~ \mathrm{where},~~~ \Psi_{01} = \psi_{01} e^{G_1(z)}.\ee
Thus, the semiclassical wavefunction executes oscillatory behaviour and therefore strongly peaked around the classical inflationary solution \eqref{sol1}, executing its viability.

\subsubsection{Hamiltonian after integrating the action by parts (Case-II):}

Here again we seek the phase-space structure of action \eqref{Anm1}, but only after controlling the total derivative terms. We therefore integrate action \eqref{Anm2} by parts, so that the total derivative terms vanish identically due to the choice, $\delta h_{ij} = 0 = \delta K_{ij}$ at the boundary. The resulting action is

\be\begin{split}\label{AMH1}
A_{11} &= \int\bigg[\Big(-\frac{3f'\dot\phi\dot{z} \sqrt z}{N} - \frac{3 f{\dot z}^2}{2 N\sqrt z} + 6 kN f \sqrt z \Big)
+ \frac{9 \beta}{\sqrt z}
\Big(\frac{{\ddot z}^2}{N^3}-\frac{2\dot N\dot z \ddot z}{N^4}+\frac{\dot N^2\dot z^2}{N^5} + \frac{2 k {\dot z}^2}{Nz}
+ 4 k^2N  \Big)\\& \hspace{2.4 in} +z^{\frac{3}{2}}\Big(\frac{1}{2N}\dot\phi^2-VN\Big)\bigg]dt.
\end{split}\ee
Thus one is left with the following point Lagrangian, where we have changed the variable $\dot z = Nx$ and introduced \big(${\dot z\over N} - x$\big) through a Lagrange multiplier $\lambda$,

\be \label{Lag1} L_{11} = -3x\sqrt z f'\dot \phi -{3N\over 2\sqrt z}fx^2 + 6Nk f\sqrt z +{9\beta\over \sqrt z}
\left({\dot x^2\over N} + {2Nkx^2\over z}  4Nk^2\right) + \left({1\over 2N}\dot\phi^2 -NV\right) z^{3\over 2}
+\lambda \left({\dot z\over N} - x\right).\ee
Following Dirac's algorithm {\footnote {We shall perform Dirac's constraint analysis explicitly in the appendix for the cases which have not been studied earlier, and leave it here to avoid repetition.}}, one arrives at

\be\label{HD11}\begin{split} H_{D2}& = N\mathcal{H}_{D2}\\& = N\left[x P_z + \frac{ \sqrt z {P_x}^2}{36\beta} +
\frac{P_{\phi}^2}{2z^{\frac{3}{2}}}+\frac{3x f'(\phi)P_{\phi}}{z} + 3f\sqrt z \Big(\frac{x^2}{2 z} - 2k \Big)
- \frac{18 k \beta}{\sqrt z} \Big(\frac{x^2}{z} + 2k\Big)+\frac{9f'^2x^2}{2\sqrt z}+Vz^{\frac{3}{2}}\right].\end{split}\ee
The same Hamiltonian was obtained earlier, in view of MHF \cite{5kMH}. It is important to note that unlike MHF, the total derivative term $\Sigma_{R_2^2}={18\beta \dot z\over N^3\sqrt z}\big(\ddot z - {\dot N\over N}\dot z\big)$, can not be removed in Dirac's formalism. Thus, once Dirac's algorithm is initiated, there is no further scope to remove the total derivative terms appearing in the action, under integration by parts. The above Hamiltonian \eqref{HD11} although looks different from \eqref{HD1}, indeed there exists a set of transformation relations in the form

\be \label{tr1} z \rightarrow z, p_z \rightarrow P_z - 18 \beta k{x\over z^{3\over 2}} + {3 f x\over 2\sqrt z};
x\rightarrow x, p_x \rightarrow P_x + {36 \beta k\over \sqrt z} + 3f\sqrt z; \phi\rightarrow \phi,
p_\phi \rightarrow P_\phi + 3 f'x\sqrt z,\ee
which relates the two Hamiltonian \eqref{HD1} and \eqref{HD11}. Now, since $p_z = P_z + {\partial F\over \partial z}, p_x = P_x + {\partial F\over \partial x}, p_\phi = P_\phi + {\partial F\over \partial \phi}$, where the generating function is, $F = 36\beta k {x\over \sqrt z} + 3f x\sqrt z$, implies a phase shift of the momenta, and therefore the transformations \eqref{tr1} are canonical. Thus, the two formalisms produce classically equivalent Hamiltonians.\\

\noindent
\textbf{canonical quantization:}\\

\noindent
However, canonical quantization is not straightforward any more, due to the coupling between $f'(\phi)$ and $P_\phi$ appearing in the fourth term of Hamiltonian \eqref{HD11}. Upon quantization one ends up with

\be \begin{split}\label{2.17}
\frac{i\hbar}{\sqrt z}\frac{\partial \Psi}{\partial z} = &-\frac{\hbar^2}{36\beta x}\left(\frac{\partial^2}{\partial x^2} + \frac{n}{x}\frac{\partial}{\partial x}
\right)\Psi -\frac{\hbar^2}{2xz^2}\frac{\partial^2 \Psi}{\partial \phi^2} + {3 \over z^{\frac{3}{2}}}\widehat{f' p_\phi}\\&\hspace{1.0 in}+ \left[\frac{3fx}{2 z} +\frac{9f'^2x}{2z}+\frac{Vz}{x} -{6kf\over x} - {18k\beta x\over z^2} - {36k^2\beta\over xz}\right]\Psi = \hat H_e\Psi,
\end{split}\ee
where, $n$ is the operator ordering index. This means, unless one knows a specific form of $f(\phi)$, the operator ordering ambiguity between $\phi$ and $P_\phi$ remains unresolved. This is the first important difference between the two Hamiltonians \eqref{HD1} and \eqref{HD11}, being noticed only in an attempt to quantize. Now in view of the form of $f(\phi)$ available from classical solution \eqref{sol1}, we have presented explicit quantum description upon Weyl ordering in \cite{5kMH} as,

\be\begin{split}\label{4.72}
i\hbar\frac{\partial \Psi}{\partial \sigma} &= -\frac{\hbar^2}{54\beta}\left(\frac{1}{x}\frac{\partial^2}{\partial x^2} + \frac{n}{x^2}\frac{\partial}{\partial x}
\right)\Psi - \frac{\hbar^2}{3x \sigma^{\frac{4}{3}}}\frac{\partial^2\Psi}{\partial\phi^2}+\frac{i\hbar}{3\sigma}\left(\frac{\phi^3+6f_1}{\phi^2}
\right)\frac{\partial\Psi}{\partial\phi}\\& \hspace{1.8 in}+\frac{i\hbar}{6\sigma}\left(\frac{\phi^3-12f_1}{\phi^3} \right)\Psi+  V_e\Psi  = \hat H_e\Psi,\\&
\mathrm{with,}~~~~~
V_e = \left[\frac{ x}{\sigma^{\frac{2}{3}}}\left(f_0+\frac{f_1}{\phi}-\frac{\phi^2}{12}\right)+\frac{3x}{\sigma^
{\frac{2}{3}}}\left(\frac{f_1}{\phi^2}+\frac{\phi}{6}\right)^2 +\frac{4f_0 \mathrm{\Lambda}^2\sigma^{\frac{2}{3}}}{x}+\frac{8f_1 \mathrm{\Lambda}^2\sigma^{\frac{2}{3}}}{x\phi}\right], \end{split}\ee
where we have substituted the form of $V(\phi)$ also in view of \eqref{sol1}. The difference between \eqref{Ds} and \eqref{4.72} is apparent. While \eqref{Ds} admits arbitrary forms of the coupling parameter $f(\phi)$ and the potential $V(\phi)$, \eqref{4.72} does not, and there is no way to translate one to the other. Further, in the process of performing semiclassical approximation, the Hamilton-Jacobi function and the zeroth order on shell action (computed from \eqref{AMH1}) have been found to match for \cite{5kMH},

\be S_0 = A_{11\mathrm{Cl}} = - 4f_0 \Lambda z^{3\over 2}+ 48\beta \Lambda^3 z^{3\over 2} - {6f_1 \Lambda\over a_0\phi_0} z^2,\ee
which is clearly different from the one obtained for the previous case (see equations \eqref{HJ1} and \eqref{HJ2} of appendix A) \footnote{Note the shuttle difference. In 2.1.1, the action to start with is \eqref{Anm2}, and so zeroth order on-shell action is computed in view of \eqref{Anm2}. On the contrary, the same is computed here from \eqref{AMH1}, which is the action to start with, i.e. after taking care of all the total derivative terms appearing in action \eqref{Anm2}.}. Upto first order of approximation, the semiclassical wavefunction was obtained as \cite{5kMH}

\be \label{PsiS} \Psi_{11} = \Psi_{011} e^{{i\over \hbar}\Lambda\left[\big(48 \beta \Lambda^2-4f_0\big) z^{3\over 2}
-{6f_1\over a_0\phi_0} z^2\right]}, ~~~ \mathrm{where},~~~ \Psi_{011} = \psi_{011} e^{G_2(z)}.\ee
Thus, although the semiclassical wavefunction here again executes oscillatory behaviour and therefore strongly peaked around the classical inflationary solution \eqref{sol1}, resulting in a viable quantum description, it is noticeable that the two wave-functions \eqref{psi1} and \eqref{PsiS} have different pre-factors and exponents. In a nutshell, although the two Hamiltonians are canonically equivalent and quantum mechanically viable, they produce different quantum dynamics. The reason is, the presence of the coupling $f(\phi)$ in the action \eqref{Anm1} has only its shear presence in the Hamiltonian \eqref{HD1} as well as in \eqref{Ds}, and doesn't affect its form from the one obtained with constant coupling, (i.e. it may be obtained simply by replacing $\alpha$ by $f(\phi)$ in equation (35) of \cite{5jMH}). Thus, for canonical quantization of the Hamiltonian obtained in the first case 2.1.1, there is no need to have specific knowledge of the coupling parameter $f(\phi)$. On the contrary, for the purpose of canonical quantization of the Hamiltonian \eqref{HD11} obtained in 2.1.2, a specific form of $f(\phi)$ is required to resolve the problem of operator ordering ambiguity appearing due to the presence of coupling between $f'(\phi)$ and $P_\phi$ in the fourth term of the above Hamiltonian \eqref{HD11}.\\

General argument runs as, it is just the change of the variables in the wave function and the phase transformation, plus the change of the integration measure, and the transformation of the momenta respecting the change of the measure, and so an unitary transformation relates the two'. Indeed the two quantum equations \eqref{Ds} and \eqref{4.72} are related by unitary transformation. Nevertheless, different forms of coupling parameter yield different quantum dynamics in case-II and so, different unitary transformations are required to relate case-I with case-II. Thus, the two quantum descriptions are not related one-to-one, and hence are distinct, and it is not clear which of the two canonically related Hamiltonians should be quantized. In the following subsection we cite yet another example.\\

\subsection{Non-minimal coupling appearing with higher order term}

In this section we take up yet another example of non-minimal coupling, to demonstrate quantum degeneracy resulting from canonically equivalent Hamiltonians. The action is chosen in the following form,

\be \label{Anm3}A_2 = \int\left[\alpha R + f(\phi) R^2 - {1\over 2}\phi_{,\mu}\phi^{,\mu} - V(\phi)\right]\sqrt{-g}~d^4x,
\ee
where the higher order term viz. $R^2$ is now coupled with the scalar field. Although, such an action might not be interesting
in the cosmological context, nevertheless, it reveals our present purpose. In the Robertson-Walker minisuperspace \eqref{RW},
the above action \eqref{Anm3} may be expressed using \eqref{R1} as

\be\label{A2}\begin{split}
A_2 = \int\bigg[{3\alpha\sqrt z}\Big(\frac{\ddot z}{ N}- \frac{\dot N \dot z}{N^2} + 2k N \Big)
+\frac{9 f(\phi)}{\sqrt z}\Big(\frac{{\ddot z}^2}{N^3} &- \frac{2 \dot N \dot z \ddot z}{N^4}
+ \frac{{\dot N}^2{\dot z}^2}{N^5} -\frac{4k\dot N \dot z}{N^2}+ \frac{4 k {\ddot z}}{N} + 4 k^2 N \Big)\\&
+z^{\frac{3}{2}}\Big(\frac{\dot\phi^2}{2N}-V N\Big)\bigg]dt.
\end{split}\ee
To obtain the field equations following variational principle, the total derivative terms, $\Sigma_R =\frac{3\alpha\sqrt z\dot z}{N},~~\text{and}~~\Sigma_{R^2} = \Sigma_{R^2_1}+ \Sigma_{R^2_2}, ~~\text{where}~~\Sigma_{R^2_1}= \frac{36 f(\phi) k\dot z}{N\sqrt z}~~\text{and}  ~~\Sigma_{R^2_2} =\frac{18 f(\phi) \dot z}{N^3\sqrt z}\left({\ddot z}-\frac{\dot z\dot N}{N} \right)$ vanish at the boundary, under the choice $\delta h_{ij}|_{\mathcal{\partial V}} = 0 = \delta K_{ij}|_{\mathcal{\partial V}}$. Thus, the scalar field and the time-time component of Einstein's field equations are,

\be \label{FE} \begin{split} &\ddot \phi + {3\over 2}{\dot z\over z}\dot \phi - {9f'(\phi)\over z^2}\left(\ddot z^2
+ 4 k \ddot z + 4 k^2\right) + V'(\phi) = 0,\\&
{3\over 8\pi G}\left({\dot z^2\over 4z^2}+{k\over z}\right) + 9f\left({\dot z\dddot z\over z^2} - {\ddot z^2\over z^2}
- {\dot z^2\ddot z\over z^3} + 6{f'\over f}\dot\phi\dot z\ddot z - 2k{\dot z^2\over z^3}  + 4{k^2\over z^2} \right)
= {1\over 2}\dot\phi^2 + V.\end{split}.\ee
The above field equations \eqref{FE} admit the following  set of inflationary solution in the flat space ($k = 0$),

\bes\label{sol} a = \sqrt z = a_0 e^{\Lambda t}; ~~\phi = \phi_0 e^{-\Lambda t};~~~
\mathrm{under ~the~ condition,}\\&~~V(\phi) = {3\Lambda^2\over 8\pi G}
+ {V_1\over \phi} + {\Lambda^2\over 2}\phi^2;~~f(\phi) = f_0 + {1\over 144\Lambda^4}\left[{V_1\over \phi}
- {\Lambda^2\over 2} \phi^2\right],\end{split}\ee
which we shall require at a later stage.

\subsubsection{Hamiltonian without integrating the action by parts (Case-I):}

Since this case has not been explored earlier, therefore we take the opportunity to compute constraint analysis in somewhat detail. In view of the new variable $x = {\dot z\over N}$, the point Lagrangian associated with action \eqref{A2} may be expressed as,

\be\label{PL} L_2 =
{3\alpha\sqrt z}\big({\dot x} + 2k N \big) +\frac{9 f(\phi)}{N\sqrt z}\big( {\dot x} + 2k N\big)^2
+z^{\frac{3}{2}}\Big(\frac{\dot\phi^2}{2N}-VN\Big) .\ee
One can easily check that the Hessian determinant vanishes and hence the above point Lagrangian \eqref{PL} is singular.
Clearly therefore Ostrogradski's technique doesn't work and it is required to follow Dirac's constraint analysis. For
this purpose, let us therefore introduce the constraint \big($\frac{\dot z}{N}-x=0$\big) through Lagrange multiplier $\lambda$
in the above point Lagrangian \eqref{PL}, which now reads as,

\be\label{D1} L_2 =
{3\alpha\sqrt z}\big({\dot x} + 2k N \big) +\frac{9 f(\phi)}{N\sqrt z}\big( {\dot x} + 2k N\big)^2
+z^{\frac{3}{2}}\Big(\frac{\dot\phi^2}{2N}-VN\Big) +\lambda\Big(\frac{\dot z}{N}-x\Big).\ee
The corresponding canonical momenta are now

\begin{equation}\label{mom}
 p_x=\frac{\partial L}{\partial\dot x}=3{\alpha\sqrt z}+\frac{18 f(\phi)}{N\sqrt{z}}\left({\dot x} + 2k N\right);~~
 p_z=\frac{\lambda}{N};~~
 p_{\phi}=\frac{z^{\frac{3}{2}}\dot\phi}{N};~~
 p_N=0;~~
 p_{\lambda}=0.
\end{equation}
The constraint Hamiltonian therefore is,

\be\begin{split}\label{D2} H_c&=\dot x p_x+\dot z p_z+\dot\phi p_{\phi}+\dot N p_N+\dot\lambda p_{\lambda}-L_2.
\end{split}\ee
We therefore require three primary constraints involving Lagrange multiplier or its conjugate viz,
$\phi_1= Np_z-{\lambda} \approx 0,\ \,  \ \phi_2=p_{\lambda} \approx 0,\ \,  \text{and},\ \ \phi_3 = p_N \approx 0$,
which are second class constraints, as $\{\phi_i,\phi_j\} \ne 0$. Note that, since the lapse function $N$ is non-dynamical,
so the associated constraint vanishes strongly, and therefore it may be safely ignored. The first two second class
constraints can now be harmlessly substituted and the modified primary Hamiltonian reads as,

\be\begin{split}\label{D3} H_{p1}&=\frac{N\sqrt z}{36 f(\phi)}p_x^2-2 k N p_x-\frac{N\alpha z p_x}{6 f(\phi)}
+\frac{N\alpha^2z^{\frac{3}{2}}}{4 f(\phi)}+\frac{Np_{\phi}^2}{2z^{\frac{3}{2}}}+VNz^{\frac{3}{2}}+\lambda x
+u_1\left(Np_z-\lambda\right)+u_2p_{\lambda}.\end{split}\ee
In the above, $u_1\ \text{and} \ u_2$ are Lagrange multipliers, and the Poisson brackets $\{x,p_x\}=\{z,p_z\}
=\{\lambda,p_{\lambda}\}=1$, hold. The requirement that the constraints must remain preserved in time is exhibited
in the Poisson brackets $\{\phi_i,H_{p1}\}$ viz,

\begin{subequations}\begin{align}&\dot\phi_1 =\{\phi_1,H_{p1}\}= -N\frac{\partial H_{p1}}{\partial z}-u_2
+ \Sigma_{i=1}^2\phi_i\{\phi_1,u_i\},\\& \dot\phi_2=\{\phi_2,H_{p1}\}=x-u_1+ \Sigma_{i=1}^2\phi_i\{\phi_2,u_i\}.
\end{align}\end{subequations}
Now, constraints must also vanish weakly in the sense of Dirac. As a result,
$\{\phi_1,H_{p1}\}=\dot\phi_1\approx 0$, requires $u_2=-N\frac{\partial H_{p1}}{\partial z}$, and
$\{\phi_2,H_{p1}\}=\dot\phi_2\approx 0,$ requires $u_1 =x$.
On thus imposing these conditions, $H_{p1}$ is then modified by the primary Hamiltonian $H_{p2}$ as,

\be\label{D4}\begin{split} H_{p2}= N\bigg[x p_z+\frac{\sqrt z p_x^2}{36 f(\phi)}-&\Big(2k+\frac{\alpha z }
{6 f(\phi)}\bigg)p_x +\frac{p_{\phi}^2}{2z^{\frac{3}{2}}}+V z^{\frac{3}{2}}+\frac{\alpha^2z^{\frac{3}{2}}}
{4f(\phi)}\bigg]\\&-N^2\Big(\frac{p_x^2}{72\sqrt z f(\phi)}-\frac{\alpha p_x}{6 f(\phi)}+\frac{3\alpha^2\sqrt z}
{8f(\phi)} -\frac{3 p_{\phi}^2}{4z^{\frac{5}{2}}}+\frac{3}{2}V \sqrt{z}\Big)p_{\lambda}.\end{split}\ee
Now, again since constraints must vanish weakly in the sense of Dirac, therefore in view of the poisson bracket
$\{\phi_1,H_{p2}\}=\dot\phi_1\approx 0$, one obtains $p_{\lambda}=0$. Thus the Hamiltonian finally takes the form,

\be\label{D5}  H_{2} = N\mathcal H_{2} =  N\left[xp_z+\frac{\sqrt z}{36 f(\phi)}p_x^2-\left(2k+\frac{\alpha z}
{6f(\phi)}\right)p_x +\frac{p_{\phi}^2}{2z^{\frac{3}{2}}}+\left(V(\phi) +\frac{\alpha^2}{4f(\phi)}\right)
z^{\frac{3}{2}}\right].\ee
It is noticeable yet again that the coupling parameter $f(\phi)$ in the action \eqref{Anm3} has only its shear presence in the Hamiltonian \eqref{D5}, and doesn't affect its form, from the one obtained with constant coupling, (i.e. it may be obtained simply by replacing $\beta$ by $f(\phi)$ in equation (35) of \cite{5jMH}). \\

\noindent
\textbf{Canonical quantization and Semiclassical wavefunction:}\\

\noindent
Under canonical quantization one ends up with (see appendix B.1)

\be\label{HS} i\hbar \frac{\partial \Psi}{\partial\sigma}=-\frac{\hbar^2}{54f(\phi)}\left(\frac{1}{x}
\frac{\partial^2}{\partial x^2} + \frac{n}{x^2}\frac{\partial}{\partial x}\right)\Psi-
\frac{\hbar^2}{3x\sigma^{\frac{4}{3}}}\frac{\partial^2\Psi}{\partial\phi^2}
+i\hbar\left(\frac{4k}{3\sigma^{\frac{1}{3}}}+\frac{\alpha\sigma^{\frac{1}{3}} }{9f(\phi)}\right)
\left(\frac{1}{x}\frac{\partial}{\partial x}-\frac{1}{2 x^2}\right) \Psi+V_e\Psi=\hat{H_e}\Psi\ee
where, the proper volume, $\sigma=z^{\frac{3}{2}}=a^3$ plays the role of internal time parameter. In the above,
$\hat H_e$ is the effective Hamiltonian operator and $V_e={2\sigma^{\frac{2}{3}}\over 3x}
\big(V+\frac{\alpha^2 }{4f(\phi)}\big)$ is the effective potential. Finally, under semiclassical approximation the Hamilton-Jacobi function is found as (see appendix B.1),

\be\label{s00} S_0 = 2 \alpha\Lambda z^{3\over 2} + 48 \Lambda^3 f_0 z^{3\over 2} - {\Lambda\over 2} a_0^2\phi_0^2 \sqrt z,\ee
which matches the zeroth order classical on-shell action computed in view of \eqref{A2} (see appendix B.1). The semiclassical wavefunction in the flat space ($k =0$) upto first order of approximation is obtained as

\be \label{SCWD}\Psi_{2} = \Psi_{02} e^{{i\over \hbar}\Lambda \big[\big(2\alpha+ 48 \Lambda^2 f_0\big) z^{3\over 2}
- {1\over 2}a_0^2\phi_0^2\sqrt z\big]}, ~~~\mathrm{where}, ~~~\Psi_{02} = \psi_{02}e^{H_1(z)}.\ee
Thus, we obtain a wavefunction which is oscillatory about classical inflationary solutions, and therefore is well behaved, establishing mathematically consistency.\\

\subsubsection{Hamiltonian after integrating the action by parts (Case-II):}

Instead, integrating the action \eqref{A2} by parts, so that some of the total derivative terms ($\Sigma_R$ and $\Sigma_{R_1^2}$) are eliminated, one ends up with the following action,

\be \label{A3} \begin{split} A_{22} = \int\bigg[\alpha\Big( - \frac{3 {\dot z}^2}{2 N \sqrt z} + 6 k N \sqrt z \Big)
+ \frac{9 f(\phi)}{\sqrt z}\Big(\frac{{\ddot z}^2}{N^3} &- \frac{2 \dot N \dot z \ddot z}{N^4} +
\frac{{\dot N}^2{\dot z}^2}{N^5} +\frac{2k {\dot z}^2}{N z} - {4k\dot z f'\dot \phi\over N f}+ 4 k^2 N \Big)\\&
+z^{\frac{3}{2}}\Big(\frac{\dot\phi^2} {2N}-VN\Big)\bigg]dt,\end{split}\ee
Now, under the change of variable $\dot z = N x$, and introducing the definition through a Lagrange multiplier, as usual, one arrives at,

\be L_{22} = \label{lag2} N\alpha\left(-{3x^2\over 2\sqrt z} + 6k\sqrt z\right) + {9f\over \sqrt z}\left({\dot x^2\over N}
+ {2kNx^2\over z} - {4kxf'\dot\phi\over f} + 4k^2 N\right) + \left({\dot\phi^2\over 2 N} - NV(\phi)\right)z^{3\over 2}
+ \lambda \left({\dot z\over N} -x\right).\ee
Proceeding as before, one ends up with the following Hamiltonian

\be\label{H22}\begin{split}& H_{22} = N \mathcal{H}_{22} = \\&N \left[x P_z + {\sqrt z P_x^2\over 36 f}  +
{P_\phi^2\over 2 z^{3\over 2}} + {36 k x f' P_\phi\over z^2} + \left({3\alpha\over 2 \sqrt z} -
{18k f \over z^{3\over 2}} +{648 k^2 f'^2 \over z^{5\over 2}} \right)x^2 - 6\alpha k \sqrt z -
{36k^2 f\over \sqrt z} + V(\phi) z^{3\over 2}\right]. \end{split}\ee
Here again we remind that in Dirac's formalism, there is no scope to remove the total derivative term $\Sigma_{R_2^2} = {18f(\phi) \dot z\over N^3\sqrt z}\big(\ddot z - {\dot N\over N}\dot z\big)$. The two Hamiltonians \eqref{D5} and \eqref{H22} are related under the following set of transformations

\be \label{tr2} z \rightarrow z,~p_z \rightarrow P_z-18 f\frac{k x }{z^{\frac{3}{2}}}+\frac{3\alpha x}{2\sqrt{z}}; ~~x
\rightarrow x,~p_x \rightarrow P_x+36 f \frac{k}{\sqrt{z}}+3 \alpha\sqrt{z};~~\phi \rightarrow \phi,~p_{\phi}
\rightarrow {P_{\phi}+{36 k f' {x\over\sqrt z}}}.\ee
Now, since $p_z = P_z + {\partial F\over \partial z}, p_x = P_x + {\partial F\over \partial x}, p_\phi = P_\phi + {\partial F\over \partial \phi}$, where the generating function is, $F = 36 f(\phi) k {x\over \sqrt z} + 3\alpha x\sqrt z$, there is only a phase shift of the momenta, and therefore the transformations \eqref{tr2} are canonical. Thus, classically the formalisms produce equivalent phase-space structures. \\

\noindent
\textbf{Canonical quantization and Semiclassical wavefunction:}\\

\noindent
The problem shoots up during canonical quantization, since, unlike the previous case (2.2.1), canonical quantization here requires primarily to fix up operator ordering ambiguity between $f'(\phi)$ and $P_\phi$ due to the coupling appearing in the third term. This is possible only after having knowledge of a specific form of $f(\phi)$, which is available in view of the classical solution \eqref{sol}. The modified Wheeler-de-Witt equation takes the look of Schr\"odinger equation (see appendix B.2), viz.,

\be\begin{split}\label{MHS} i\hbar \frac{\partial \Psi}{\partial\sigma}=&-\frac{\hbar^2}{54f(\phi)}
\left(\frac{1}{x}\frac{\partial^2}{\partial x^2} + \frac{n}{x^2}\frac{\partial}{\partial x}\right)\Psi
-\frac{\hbar^2}{3x\sigma^{\frac{4}{3}}}\frac{\partial^2\Psi}{\partial\phi^2}\\&
+ i\hbar{k\over 4\Lambda^4 \sigma^{5\over 3}}\left[\left({V_1\over \phi^2}
+ \Lambda^2\phi\right){\partial\Psi\over\partial\phi} + \left({\Lambda^2\phi^3
- 2V_1\over 2\phi^3}\right)\Psi\right] +V_e\Psi=\hat{H_e}\Psi\\&
\mathrm{where},~~V_e = \left({3\alpha\over 2 \sigma^{2\over 3}} - {18k f \over \sigma{^4\over 3}}
+{648 k^2 f'^2 \over \sigma^2} \right)x  - {36k^2 f\over x \sigma^{2\over 3}} + {\sigma^{2\over 3}\over x}\big(V(\phi)
-6\alpha k\big).
\end{split}\ee
In the above, the proper volume, $\sigma=z^{\frac{3}{2}}=a^3$ plays the role of internal time parameter, and we have
performed the Weyl symmetric operator ordering between $\hat f'(\phi)$ and $\hat P_\phi$. Further, $\hat H_e$ and $V_e$
are the effective Hamiltonian operator and the effective potential respectively. One can clearly notice considerable difference between the effective Hamiltonian appearing in \eqref{HS} and \eqref{H22}. As before, the effective Hamiltonian appearing in \eqref{HS} admits arbitrary forms of the coupling parameter $f(\phi)$ and the potential $V(\phi)$, the same appearing in \eqref{H22} applies only for a specific forms of the same. More clearly, the quantum counterpart of the Hamiltonian \eqref{H22} is different for different forms of $f(\phi)$ and $V(\phi)$. Further, under semiclassical approximation the Hamilton-Jacobi function reads as (see appendix B.2),

\be \label{S0m} S_0 = - 4\alpha\Lambda z^{3\over 2} +48 \Lambda^3 f_0 z^{3\over 2}
- {\Lambda\over 2} a_0^2\phi_0^2\sqrt z .\ee
Therefore, the semiclassical wavefunction upto first order approximation is,

\be \label{scwd}\Psi_{2M} = \Psi_{022} e^{{i\over \hbar}\Lambda \big[\big(-4\alpha+ 48 \Lambda^2 f_0\big) z^{3\over 2}
- {1\over 2}a_0^2\phi_0^2\sqrt z\big]}, ~~~\mathrm{where}, ~~~\Psi_{022} = \psi_{022}e^{H_2(z)}.\ee
The wavefunction here again executes oscillatory behaviour about classical inflationary solutions, however, the wave functions \eqref{SCWD} and \eqref{scwd} are different both in the prefactors as well as in the exponents. The two quantum descriptions are related under unitary transformation is the same general argument. But, different forms of coupling parameter yields different quantum dynamics in case-II, and therefore case-I can not be related to case-II through a unique unitary transformation. This example therefore again expatiates quantum degeneracy with canonically equivalent Hamiltonians. In the following we cite yet another example, which would finally lead us to choose a unique quantum description of a classical theory.

\subsection{Modified Einstein-Gauss-Bonnet-Dilatonic coupled action.}

Our final example is the `Einstein-Gauss-Bonnet-Dilatonic coupled action in the presence of scalar curvature invariant term' (MEGBD), which has been treated in a recent article \cite{5jMH}. The action was taken in the following form,

\be\label{Agb} A_{gb} =\int\sqrt{-g}\;d^4x\left[\alpha{R}+\beta R^2+\gamma(\phi)\mathcal{G}
-\frac{1}{2}\phi_{,\mu}\phi^{,\nu}-V(\phi)\right].\ee
In the above action \eqref{Agb} $\gamma(\phi)$ is the coupling parameter and $V(\phi)$ is the dilatonic potential. The expression for the Gauss-Bonnet term in the Robertson-Walker minisuperspace \eqref{RW} reads as,

\be \label{G} \mathcal{G} = R^2 - 4 R_{\mu\nu}R^{\mu\nu} + R_{\alpha\beta\gamma\delta}R^{\alpha\beta\gamma\delta} =\frac{24}{N^3 a^3}\Big(N\ddot a - \dot N \dot a\Big)\Big(\frac{\dot a^2}{N^2} + k\Big) =
{12\over N^2}\Big({\ddot z\over z}-{1\over 2}{\dot z^2\over z^2}-{\dot N\over N}{\dot z\over z}\Big)
\Big({1\over 4 N^2}{\dot z^2\over z^2} + {k\over z}\Big)\ee
as a result, the action (\ref{Agb}) in terms of the basic variable $z$ takes the form,
\be\begin{split}\label{A12}
A_{gb} &= \int\Bigg[{3\alpha\sqrt z}\Big(\frac{\ddot z}{ N}- \frac{\dot N \dot z}{N^2} + 2k N \Big)+\frac{9\beta}
{\sqrt z}\Big(\frac{{\ddot z}^2}{N^3} - \frac{2 \dot N \dot z \ddot z}{N^4} + \frac{{\dot N}^2{\dot z}^2}{N^5}
-\frac{4k\dot N \dot z}{N^2}+ \frac{4 k {\ddot z}}{N} + 4 k^2 N\Big) \\& \hspace{0.5 in}+ \frac {3 \gamma (\phi)}
{N \sqrt z}\Big(\frac{{\dot z}^2 \ddot z}{N^2 z} + 4 k \ddot z - \frac{{\dot z}^4}{ 2 N^2 z^2}
- \frac{\dot N {\dot z}^3}{N^3 z} - \frac{2 k {\dot z}^2}{z}- \frac{4 k \dot N \dot z}{N}\Big)+z^{\frac{3}{2}}
\Big(\frac{1}{2N}\dot\phi^2-VN\Big)\Bigg]dt.
\end{split}\ee
The variation of the above action \eqref{A12} leaves three total derivative terms, viz. $\Sigma_R =\frac{3\alpha\sqrt z\dot z}{N},~~\Sigma_{R^2} = \Sigma_{R^2_1}+ \Sigma_{R^2_2}$, where $\Sigma_{R^2_1}= \frac{36\beta k\dot z}{N\sqrt z} ~~\text{and}  ~~\Sigma_{R^2_2} =\frac{18 \beta\dot z}{N^3\sqrt z}\left({\ddot z}-\frac{\dot z\dot N}{N} \right)$ and $\Sigma_\mathcal{G} = \gamma(\phi)\frac {\dot z}{N \sqrt z}\Big(\frac{{\dot z}^2}{N^2 z} + 12 k\Big)$. The $\phi$ variation equation and the $(^0_0)$ component of the Einstein's field equation in terms of the scale factor
are,

\be\label{phi} 24\left({\dot a^2\ddot a\over N^3} + k{\ddot a\over N} - {\dot a^3\dot N\over N^4} -
{k\dot a\dot N\over N^2}\right)\gamma' - {a^3\over N}\left(\ddot \phi + 3 {\dot a\over a}\dot\phi+N^2 V' -
{\dot N\over N}\dot\phi\right) = 0.\ee
\be\begin{split}\label{e00}& \frac{6\alpha}{a^2}\Big(\frac{\dot a^2}{N^2}+k\Big) + \frac{36\beta}{a^2N^4}
\Bigg(2\dot a\dddot a-2\dot a^2\frac{\ddot N}{N}-\ddot a^2-4\dot a\ddot a\frac{\dot N}{N}+2\dot a^2
\frac{\ddot a}{a}+5\dot a^2\frac{\dot N^2}{N^2}
-2\frac{\dot a^3\dot N}{a N}\\&\hspace{0.8 in}-3\frac{\dot a^4}{a^2}-2kN^2\frac{\dot a^2}{a^2}+\frac{k^2N^4}{a^2}\Bigg)
+ \frac{24\gamma'\dot a\dot\phi}{N^2a^3}\Big(\frac{\dot a^2}{N^2}+k\Big) - \Big(\frac{\dot\phi^2}{2N^2}+V\Big)=0,
\end{split}\ee
In the above, prime denotes derivative with respect to $\phi$. Note that, Gauss-Bonnet being topologically invariant is merely a total derivative term in four-dimension. So, if the dilatonic coupling parameter $\gamma$ somehow becomes constant, one can at once see from the above field equations that there would be no contribution to the field equations, and hence to the Hamiltonian. Now, the above set of equations admits the following inflationary solutions ($k = 0, N= 1$)

\be\begin{split}\label{111}&a=a_0e^{\Lambda t}~~\text{and} ~~\phi=\phi_0 e^{-\Lambda t},~~~
\mathrm{under ~the~ condition,}\\&\gamma = -{\phi^2\over 48 \Lambda^2}, \;\;
V = {6\alpha\Lambda^2} + {1\over 2}\Lambda^2\phi^2. \end{split}\ee
In the following subsections, we again briefly demonstrate how two canonically equivalent Hamiltonians, produce different quantum dynamics, which are not related any further.

\subsubsection{Hamiltonian without integrating the action by parts (Case-I):}

The Hamiltonian following Dirac's constrained analysis, without taking care of the total derivative terms a-priori was found in \cite{5jMH} as,

\be\begin{split}\label{Hgbd} H_{3D}&=N{\mathcal{H}_{3D}} = N\Bigg[x p_z+\frac{\sqrt{z}p_x^2}{36\beta}-
\left(\frac{\alpha z }{6\beta}+\frac{\gamma x^2 }{6\beta z}+\frac{2k\gamma}{3\beta}+2k\right)p_x+
\frac{p_{\phi}^2}{2z^{\frac{3}{2}}}+\left(\frac{\gamma^2}{4\beta z^{\frac{5}{2}}} +
\frac{3\gamma}{2z^{\frac{5}{2}}}\right) x^4\\& \hspace{0.7 in}+\frac{\alpha^2 z^{\frac{3}{2}}}{4\beta}+
\left(\frac{\alpha \gamma}{2\beta \sqrt{z}}+\frac{12k\gamma}{z^{\frac{3}{2}}} +\frac{2k\gamma^2}
{\beta z^{\frac{3}{2}}}\right)x^2+\frac{2\alpha k\gamma\sqrt{z}}{\beta}+\frac{24 k^2\gamma}
{\sqrt z}+\frac{4k^2\gamma^2}{\beta \sqrt{z}} +Vz^{\frac{3}{2}}\Bigg].\end{split}\ee
The Hamiltonian \eqref{Hgbd} produces correct classical field equations, and therefore is classically viable. But, one can observe an immediate consequence of not integrating the action by parts. Gauss-Bonnet term contributes to the Hamiltonian even if $\gamma$ is a constant. Under canonical quantization, using Weyl symmetric ordering, one arrives at the following Schrodinger-like equation \cite{5jMH}, viz.,

\be\begin{split}\label{q12}
i\hbar\frac{\partial \Psi}{\partial \sigma} =& -\frac{\hbar^2}{54\beta x}\left(\frac{\partial^2}{\partial x^2} + \frac{n}{x}\frac{\partial}{\partial x}\right)\Psi+i\hbar\left(\frac{4k}{3\sigma^{\frac{1}{3}}}+\frac{\alpha \sigma^{\frac{1}{3}}}{9\beta}+\frac{4k\gamma(\phi)}{9\beta \sigma^{\frac{1}{3}}}\right)\left(\frac{1}{x}\frac{\partial\Psi}{\partial x}-\frac{\Psi}{2x^2}\right)\\&+\frac{i\hbar\gamma(\phi)}{18\beta \sigma}\left(2x\frac{\partial\Psi}{\partial x}+\Psi\right)-\frac{\hbar^2}{3x\sigma^{\frac{4}{3}}}\frac{\partial^2\Psi}{\partial \phi^2}+V_e\Psi=\hat{H_e}\Psi\\&
V_e=\frac{\alpha^2\sigma^{\frac{2}{3}}}{6\beta x}+  \Big(\frac{x^2}{\sigma}+\frac{4k}{\sigma^{\frac{1}{3}}}\Big)^2{\gamma^2\over 6\beta x}+\Big(\frac{x^3}{\sigma^{\frac{4}{3}}}+\frac{\alpha x }{3\beta {\sigma^{\frac{2}{3}}}}+\frac{4k\alpha}{3\beta x}+\frac{8kx}{\sigma^{\frac{2}{3}}}+\frac{16k^2}{x\sigma^{\frac{2}{3}}} \Big)\gamma +\frac{2 V\sigma^{\frac{2}{3}}}{3 x}.
\end{split}\ee
Where $\sigma=z^{\frac{3}{2}}=a^3$ plays the role of internal time parameter, as before. Quantum mechanical probability interpretation holds for the above Schrodinger-like equation \eqref{q12} for $n = -1$, and the Hamilton-Jacobi function has been evaluated as \cite{5jMH}

\be\label{S13} S_0=2\alpha \mathrm{\Lambda} z^{\frac{3}{2}}+48\beta \mathrm{\Lambda}^3 z^{\frac{3}{2}}-{\mathrm{\Lambda}\over 2}
a_0^2\phi_0^2\sqrt z = A_{cl},\ee
and therefore the semiclassical wave-function up-to first order approximation reads as,

\be \label{psi3}\Psi_{3} = \Psi_{03} e^{{i\over \hbar} \big[2\alpha \mathrm{\Lambda} z^{\frac{3}{2}}
+48\beta \mathrm{\Lambda}^3 z^{\frac{3}{2}}-{\mathrm{\Lambda}\over 2} a_0^2\phi_0^2\sqrt z\big]}, ~~~\mathrm{where},
~~~\Psi_{03} = \psi_{03}e^{J_1(z)}.\ee
The wavefunction here again executes oscillatory behaviour about classical inflationary solutions \eqref{111}, and therefore is perfectly well behaved.\\

\subsubsection{Hamiltonian after integrating the action by parts (Case-II):}

After integrating the action \eqref{A12} by parts, so that all total derivative terms but $\Sigma_{R^2_2}$ are removed, one is left with the following action,

\be\begin{split}\label{A20}
A_{33} = \int\bigg[\alpha\left( - \frac{3 {\dot z}^2}{2 N \sqrt z} + 6 k N \sqrt z \right) & + \frac{9 \beta}{\sqrt z}\left(\frac{{\ddot z}^2}{N^3} - \frac{2 \dot N \dot z \ddot z}{N^4} + \frac{{\dot N}^2{\dot z}^2}{N^5} + \frac{2 k {\dot z}^2}{N z} + 4 k^2 N \right) \\&- \frac {\gamma^{'} \dot z \dot {\phi}}{N \sqrt z}\left(\frac{{\dot z}^2}{N^2 z} + 12 k\right) +z^{\frac{3}{2}}\left(\frac{1}{2N}\dot\phi^2-VN\right)\bigg]dt.
\end{split}\ee
To initiate Dirac's algorithm, as before, it is required to change the variable $\dot z = N x$ in \eqref{A20}, and insert it through a Lagrange multiplier in the point Lagrangian as before, to obtain

\be \begin{split}L_{33} = &{9\beta\over N\sqrt z}\left(\dot x^2 + {2k N^2 x^2\over z} + 4k^2 N^2\right) -
{x\gamma'\dot \phi\over \sqrt z}\left({x^2\over z}+12 k\right) + z^{3\over 2}\left({\dot\phi^2\over 2N}-NV\right)
\\&\hspace{2.2 in}-6\alpha N \left({x^2\over 4\sqrt z}-k\sqrt z\right) + \lambda \left({\dot z\over N}-x\right).\end{split}\ee
Canonical momenta are,

\be p_x = {18\beta\over N\sqrt z} \dot x;\;\;\;\;p_\phi = {z^{3\over 2}\dot\phi\over N} - {\gamma'x\over \sqrt z}
\left({x^2\over z} + 12 k \right);\;\;\;\;p_z = {\lambda\over N};\;\;\;\;p_\lambda = 0 = p_N.\ee
Hence,

\be\begin{split} H_{p1} &= {N\sqrt z\over 18\beta}p_x^2 + {N p_\phi^2\over z^{3\over 2}} + {N x\gamma'\over z^2}
\left({x^2\over z}+12k\right)p_\phi + \dot z{\lambda\over N} -{N\sqrt z\over 36\beta}p_x^2 -{18N\beta k\over \sqrt z}
\left({x^2\over z} + 2 k\right) + 6\alpha N \left({x^2\over 4\sqrt z}-k\sqrt z\right)\\&
+ {N x\gamma'\over \sqrt z}\left({x^2\over z}+12k\right)\left[{p_\phi\over z^{3\over 2}} +{x \gamma'\over z^2}
\left({x^2\over z}+12k\right)\right] + 6\alpha N \left({x^2\over 4\sqrt z}-k\sqrt z\right)\\&
-{N^2 z^{3\over 2}\over 2 N}\left[{p_\phi^2 \over z^3} + {2 x\gamma'\over z^{7\over 2}}\left({x^2\over z}+12k\right)
p_\phi + {x^2 \gamma'^2 \over z^4}\left({x^2\over z} + 12 k\right)^2\right] \\&+ z^{3\over 2} N V - \dot z{\lambda\over N}
+ \lambda x + u_1(Np_x - \lambda) + u_2 p_\lambda\\&
=N\left[{\sqrt z\over 36\beta}p_x^2 + {p_\phi^2\over 2 z^{3\over 2}} + 6\alpha N \left({x^2\over 4\sqrt z}-k\sqrt z\right)+ {x\gamma'\over z^2}
\left({x^2\over z}+12k\right)p_\phi + {x^2 \gamma'^2\over 2 z^{5\over 2}}\left({x^2\over z}+12k\right)^2
- {18 k \beta \over \sqrt z}\left({x^2\over z} + 2k\right)+ z^{3\over 2}  V\right] \\&+ \lambda x + u_1(Np_x - \lambda)
+ u_2 p_\lambda,~~~\mathrm{where},~~~\phi_1 = Np_z - \lambda \approx 0,~~~\phi_2 = p_\lambda \approx 0.\end{split}\ee
Now, since

\be\begin{split} & \dot\phi_1  = \{\phi_1, H_{p1}\} = -N{\partial H_{p1}\over\partial z} - u_2 \approx 0;\Rightarrow
u_2 =  -N{\partial H_{p1}\over\partial z},\\&\dot\phi_2 = \{\phi_2, H_{p1}\} = -x + u_1 \approx 0;\Rightarrow u_1 = x,
\end{split}\ee
hence the primary Hamiltonian is modified to

\be\begin{split} H_{p2} &= N\Bigg[x p_z + {\sqrt z\over 36\beta}p_x^2 + {p_\phi^2\over 2 z^{3\over 2}} +
{x\gamma'\over z^2}\left({x^2\over z}+12k\right)p_\phi + {x^2 \gamma'^2\over 2 z^{5\over 2}}\left({x^2\over z}
+12k\right)^2 \\&\hspace{1.0 in}+ 6\alpha  \left({x^2\over 4\sqrt z}-k\sqrt z\right) - {18 k \beta \over \sqrt z}\left({x^2\over z} + 2k\right)+ z^{3\over 2}  V\Bigg]
- N p_\lambda {\partial H_{p1}\over\partial z}.\end{split}\ee
Now one can check that $\dot \phi_1 = \{\phi_1, H_{p2}\} \approx 0,~\Rightarrow p_\lambda = 0$, and $\dot\phi_1$
trivially vanishes. Therefore, the Hamiltonian finally reads as,

\be\begin{split} \label{Hgbm} H_{33}& = N \mathcal{H}_{33} = N\Bigg[x p_z + {\sqrt z\over 36\beta}p_x^2 + {p_\phi^2\over 2 z^{3\over 2}} +
{x\gamma'\over z^2}\left({x^2\over z}+12k\right)p_\phi + {x^2 \gamma'^2\over 2 z^{5\over 2}}\left({x^2\over z}
+12k\right)^2 \\&\hspace{1.8 in}+ 6\alpha \left({x^2\over 4\sqrt z}-k\sqrt z\right) - {18 k \beta \over \sqrt z}\left({x^2\over z} + 2k\right)+ z^{3\over 2}  V\Bigg].
\end{split}\ee
Again the two Hamiltonians \eqref{Hgbd} and \eqref{Hgbm} are related via the transformation relations

\be\begin{split} \label{tr3}& z \rightarrow z,~p_z \rightarrow P_z-18 \beta\frac{k x }{z^{\frac{3}{2}}}+\frac{3\alpha x}{2\sqrt{z}}-{6k\gamma x\over z^{\frac{3}{2}}} - {3\gamma x^3\over 2 z^{\frac{5}{2}}}; \\&
x\rightarrow x,~p_x \rightarrow P_x+36 \beta \frac{k}{\sqrt{z}}+3 \alpha\sqrt{z} +{12k\gamma\over\sqrt z} + {3\gamma x^2\over {z^{\frac{3}{2}}}};\\&
\phi \rightarrow \phi,~p_{\phi}\rightarrow P_{\phi}+\left({x^3\over z^{\frac{3}{2}}}+ {12 kx\over\sqrt z}\right)\gamma'.\end{split}\ee
Now, as before, since $p_z = P_z + {\partial F\over \partial z}, p_x = P_x + {\partial F\over \partial x}, p_\phi = P_\phi + {\partial F\over \partial \phi}$, where the generating function is, $F = {36 \beta kx\over \sqrt z} + 3\alpha x\sqrt z + {12k\gamma x\over \sqrt z} + {\gamma x^3\over z^{3\over 2}}$, there is only a phase shift of the momenta, and therefore the set of transformations \eqref{tr3} is canonical. As before, although there is no contradiction at the classical level, canonical quantization unveils appreciable difference between the two Hamiltonians \eqref{Hgbd} and \eqref{Hgbm}. For example, while the Hamiltonian \eqref{Hgbd} contains the coupling parameter $\gamma(\phi)$ every now and then, the Hamiltonian \eqref{Hgbm} involves only the derivative ($\gamma'$) of the same. The Schr\"odinger equation reads as,

\be\begin{split}\label{5}
i\hbar\frac{\partial \Psi}{\partial \sigma} &= -\frac{\hbar^2}{54\beta}\left(\frac{1}{x}\frac{\partial^2}{\partial x^2} + \frac{n}{x^2}\frac{\partial}{\partial x}\right)\Psi - \frac{\hbar^2}{3x \sigma^{\frac{4}{3}}}\frac{\partial^2\Psi}{\partial \phi^2} + {2\over 3}\left({x^2\over \sigma^{7\over 3}} + {12 k\over \sigma^{5\over 3}}\right)\widehat{\gamma' P_\phi} + V_e\Psi = H_e\Psi\\&
V_e = \left({x^2\over \sigma}+{12 k\over \sigma^{1\over 3}}\right)^2{\gamma'^2 x\over 3\sigma^{4\over 3}}+{\alpha\sigma^{1\over 3}\over x}\left({x^2\over \sigma} - {4k\over\sigma^{1\over 3}}\right) - {12 k\beta\over x \sigma^{1\over 3}}\left({x^2\over \sigma}+{2k\over \sigma^{1\over 3}}\right) + {2 V\sigma^{2\over 3}\over 3 x},\end{split}\ee
where, $\sigma = z^{3\over 2}$ plays the role of internal time parameter, as usual. Clearly, we need a form of $\gamma(\phi)$ to resolve the operator ordering ambiguity appearing in the third time on the right hand side. Thus the two quantum equations \eqref{q12} and \eqref{5} suffer from operator ordering ambiguities in different variables, and such ambiguity cannot be removed from \eqref{Hgbm} unless, the form of the coupling parameter $\gamma(\phi)$ is known a-priori. This, again depicts important difference between the two. In view of the form available from the classical solution \eqref{111}, one may express the above Schr\"odinger equation as,

\be\begin{split}\label{6}
i\hbar\frac{\partial \Psi}{\partial \sigma} &= -\frac{\hbar^2}{54\beta}\left(\frac{1}{x}\frac{\partial^2}{\partial x^2} + \frac{n}{x^2}\frac{\partial}{\partial x}\right)\Psi - \frac{\hbar^2}{3x \sigma^{\frac{4}{3}}}\frac{\partial^2\Psi}{\partial \phi^2} + {1\over 36 \Lambda^2}\left({x^2\over \sigma^{7\over 3}} + {12 k\over \sigma^{5\over 3}}\right)\left(2\phi{\partial \Psi\over \partial\phi} + \Psi\right) + V_e\Psi = H_e\Psi.\end{split}\ee
The effective Hamiltonian is hermitian, as before under the choice $n = -1$, and probability interpretation holds as well. The Hamiltonians \eqref{Hgbm} also passes all the tests in the quantum domain, but the difference between the two becomes more prominent while comparing the semi-classical wave-functions \cite{5jMH}. The Hamilton-Jacobi function hereto matches with the zeroth-order on-shell actions (computed in view of action \eqref{A20}), viz.

\be\label{S33} S_0= - 4\alpha \mathrm{\Lambda} z^{\frac{3}{2}}+48\beta \mathrm{\Lambda}^3 z^{\frac{3}{2}}
-{\mathrm{\Lambda}\over 3} a_0^2\phi_0^2\sqrt z = A_{cl},\ee
which is again different from the one \eqref{S13} obtained in the previous sub-subsection. The semiclassical wave-function up-to first order approximation reads as,

\be \label{psi33}\Psi_{33} = \Psi_{033} e^{{i\over \hbar} \big[-4\alpha \mathrm{\Lambda} z^{\frac{3}{2}}
+48\beta \mathrm{\Lambda}^3 z^{\frac{3}{2}}-{\mathrm{\Lambda}\over 3} a_0^2\phi_0^2\sqrt z\big]}, ~~~\mathrm{where},
~~~\Psi_{033} = \psi_{033}e^{J_2(z)}.\ee
The wavefunction here again executes oscillatory behaviour about classical inflationary solutions \eqref{111}, and therefore is also perfectly well behaved. One can clearly notice that the semi-classical wavefunctions \eqref{psi3} and \eqref{psi33} contain different pre-factors and also different exponents and so the two Hamiltonians \eqref{Hgbd} and \eqref{Hgbm} present different quantum descriptions altogether, resulting in the so-called degenerate Hamiltonian operator.\\

\section{Probing deep into the problem:}

Although it appears to be a run of the machine, however, in the process of exploring different cases, we have already probed deep into the problem. It's true that total derivative terms present in an action, do not affect the equations of motion, rather it just change the canonical momenta in a way that only amounts to a canonical transformation. But, the problem is, two canonically equivalent Hamiltonians (derived from an action differing by total derivative terms) lead to inequivalent quantum descriptions. If total derivative terms are not taken care of, the coupling parameter does not affect quantum dynamics except its shear presence. On the contrary, if total derivative terms are taken care of, then quantum description depends inherently on its typical form. Thus, our case-I is a unique description of the classical theory, while case-II yields different quantum dynamics depending on the form of the coupling parameter. Therefore, as already stated, the two cases are not related one-to-one. This means, quantum mechanically, the Hamiltonian operators are not equivalent, and as a result, different energy eigenvalues would emerge. Further, Hamilton-Jacobi functions are different in the two cases, which is uncanny. Finally, different prefactors of the semiclassical wavefunctions lead to different normalization conditions, with different probability densities. Thus, we need to choose a unique Hamiltonian for the purpose of quantization.\\

\noindent
We can find a route to our destination, if we concentrate on the case studied last in subsection (2.3). First, we remember that Gauss-Bonnet term \eqref{G} appearing in the action \eqref{Agb} is topologically invariant in four dimensions and acts as a total derivative term, which does not contribute to classical field equations. This is the reason for associating it with a dilatonic coupling parameter $\gamma(\phi)$. Note that, for a constant coupling parameter, $\gamma'(\phi) = 0$, the Hamiltonian \eqref{Hgbm} does not involve any trace of Gauss-Bonnet term, and reduces to the one, derived in view of the action $A = \int [\alpha R + \beta R^2 -{1\over 2}\phi_{,\mu}\phi^{,\mu} - V(\phi)]\sqrt{-g} d^4 x$, earlier \cite{5jMH}. Of-course, this is expected and quite natural. Unfortunately, for $\gamma =$ constant, the Gauss-Bonnet term remains very much present in the Hamiltonian \eqref{Hgbd} and its quantum counterpart \eqref{q12}, which is definitely illogical and mathematically inconsistent. Thus, although, the two Hamiltonians \eqref{Hgbd} and \eqref{Hgbm} are canonically equivalent, \eqref{Hgbd} clearly leads to a wrong quantum description of the classical theory under consideration, and therefore not suitable for canonical quantization. To avoid such unwanted feature, it is therefore mandatory to take care of the total derivative terms under integrating the action by parts, a-priori, in general.\\

Another mathematical inconsistency is somewhat hidden in Case-I of Dirac's formalism, i.e. if the action is not made free from the total derivative terms. It is true that Dirac's formalism has so far been hailed with great success to handle constrained system, and so it's dreary to think that the formalism is less propitious in higher-order theories of gravity. Still such hidden  inconsistency is revealed while Horowitz' formalism (HF) is initiated for canonical formulation of higher-order theory of gravity. In fact MEGBD theory depicts such inconsistency, if total derivative terms are not taken care of a-priori, as will be expatiated following HF in the following subsection.\\

In the following subsections we again follow two different routes towards canonical formulation of higher order theory, which are known by the name Horowitz' formalism (HF) and modified Horowitz' formalism (MHF), as mentioned in the introduction. In HF, an auxiliary variable is chosen by varying the action with respect to the highest derivative of the field variable present, and then it is introduced in the action judiciously, so that the action is canonical. The total derivative terms are then removed under integration by parts, and the Hamiltonian so obtained is finally expressed in terms of the basic variables ($h_{ij},~K_{ij}$). In MHF, on the contrary, the total derivative terms $\Sigma_R$ and $\Sigma_{R^2_1}$ (and also $\Sigma_\mathcal{G}$, when Gauss-Bonnet term is present) are eliminated in the first step, and then auxiliary variable is invoked, which is then entered into the action. The rest of the total derivative terms, viz. $\Sigma_{R^2_2}$ is removed at this stage, which is the reason for splitting $\Sigma_{R^2}$. The Hamiltonian so obtained is then expressed in terms of the basic variables ($h_{ij},~K_{ij}$), as before. Thus in HF, all the total derivative terms are eliminated in a single stage, while in MHF, the boundary terms are eliminated in two steps. The question naturally arises, if these total derivative terms match the surface terms obtained under variation of the action? This is what we exactly pose in the following subsections.

\subsection{Scalar-tensor theory of gravity in the presence of higher-order term:}

\subsubsection{Horowitz' Formalism (HF):}

As mentioned, in the HF, it is customary to define the auxiliary variable, without handling the total derivative terms present in the action. Therefore, in view of {\eqref{Anm2}} one finds,

\be\label{Q1} Q={\partial A_{1}\over \partial\ddot z} =\frac{3f\sqrt{z}}{N}+ \frac{18\beta}{N\sqrt{z}}\Big(\frac{\ddot z}{N^2}-
\frac{\dot N\dot z}{N^3}+2k\Big),\ee
which is then introduced in the action, so that

\be\begin{split}\label{1}
A_{1} &= \int\Big[Q\ddot z-\frac{N^3\sqrt{z}Q^2}{36\beta}-\frac{Nf^2 z^{\frac{3}{2}}}{4\beta}+\frac{fN^2 Qz}{6\beta}+2kQN^2 -\frac{\dot N\dot z Q}{N} + \frac{z^{\frac{3}{2}}\dot\phi^2}{2N}-VNz^{\frac{3}{2}}\Big]dt.\end{split} \ee
After integration by parts, resulting canonical action reads

\be\begin{split}\label{12}
A_{1} &= \int\Big[-\dot Q\dot z-\frac{N^3\sqrt{z}Q^2}{36\beta}-\frac{Nf^2 z^{\frac{3}{2}}}{4\beta}+\frac{fN^2 Qz}{6\beta}+2kQN^2 -\frac{\dot N\dot z Q}{N} + \frac{z^{\frac{3}{2}}\dot\phi^2}{2N}-VNz^{\frac{3}{2}} \Big]dt.\end{split} \ee
It is important to note that the total derivative term

\be\label{TDT1} Q\dot z = {3f\sqrt z\dot z\over N} + {36\beta\sqrt z\dot z\over N^3}\left({\ddot z\over 2z} + {N^2 k\over z} - {\dot N \dot z\over 2Nz}\right) = \Sigma_R + \Sigma_{R^2},\ee
is exactly the same total derivative term that appears under variation of the action \eqref{Anm2}. Now the canonical momenta are,

\be\label{ph} p_z =-\dot Q-{\dot N Q\over N};\;\;\; p_\phi = {{z^{\frac{3}{2}}\dot\phi}\over N};\;\;\; p_Q =
  -\dot z;\;\;\;p_N=-\frac{\dot z Q}{N},
\ee
Despite, the presence of constraint in the form $Qp_Q - Np_N = 0$, the Hamiltonian in phase space variables is readily obtained without performing constraint analysis as (see appendix B.3 of \cite{5kMH}),

\be\begin{split}\label{hh1}
 H_1 &=-p_Q p_z+\frac{N^3\sqrt{z}Q^2}{36\beta}-2kQN^2-\frac{fN^2 Qz}{6\beta}+\frac{Nf^2 z^{\frac{3}{2}}}{4\beta} +\frac{Np^2_{\phi}}{2z^{\frac{3}{2}}}
+VNz^{\frac{3}{2}}.\end{split}\ee
It is now required to express the Hamiltonian in terms of basic variables ($z$ and $x =\frac{\dot z}{N}$), instead of ($z$ and $Q$).
Since, $p_Q =-\dot z=-Nx$ and $Q =\frac{p_x}{N}$, one therefore is required to make the following canonical transformation under the replacements, $p_Q$
by $-Nx$ and $Q$ by $\frac{p_x}{N}$ in the Hamiltonian (\ref{hh1}), which therefore finally results in,

\be\begin{split}\label{HH1}
 H_1 & = N\mathcal{H}_{1} = N\left[x p_z + {\sqrt z\over 36\beta} p_x^2 -\left({f(\phi) z\over 6\beta} + 2k\right)p_x  +  {p_\phi^2\over 2z^{3\over 2}} + {f^2 z^{3\over 2}\over 4\beta} + V z^{3\over 2}\right],\end{split}\ee
and diffeomorphic invariance is established. At a first glance it might appear that the transformation is not canonical, since $\{Q, p_N\} = \{{p_x\over N}, p_N\} = - {p_x\over N^2}$, and  $\{p_Q, p_N\} = \{-Nx, p_N\} = -x$, do not vanish. However, $N$ being a Lagrange multiplier, should not be treated as a basic variable, as may be expatiated starting from a more fundamental variable, $q = NQ = 3f\sqrt{z}+ \frac{18B}{\sqrt{z}}\Big(\frac{\ddot z}{N^2}-\frac{\dot N\dot z}{N^3}+2k\Big)$, in view of \eqref{Q1}. In that case, the action \eqref{12} reads as,

\bes A_{1} &= \int\Big[-{\dot q\dot z\over N}-\frac{N\sqrt{z}q^2}{36\beta}-\frac{Nf^2 z^{\frac{3}{2}}}{4\beta}+\frac{N f q z}{6\beta}+2 N k q + \frac{z^{\frac{3}{2}}\dot\phi^2}{2N}-VNz^{\frac{3}{2}} \Big]dt.\end{split} \ee
One can notice that $\dot N$ term does not appear in the above action any more, as it should be. One can therefore, cast the Hamiltonian in view of the canonical momenta

\be p_z = -{\dot q\over N},~~~p_q = -{\dot z\over N},~~p_\phi = z^{3\over 2}{\dot \phi\over N}\ee
in a straightforward manner as

\be\begin{split}
 H_1 &=N\left[-p_q p_z + \frac{\sqrt{z}q^2}{36\beta}-2k q -\frac{f q z}{6\beta}+\frac{f^2 z^{\frac{3}{2}}}{4\beta} +\frac{p^2_{\phi}}{2z^{\frac{3}{2}}}+Vz^{\frac{3}{2}}\right].\end{split}\ee
Now, one can easily observe that $N$ should not be treated as variable, since it acts only as a Lagrange multiplier. Therefore, it is allowed to make the canonical transformations $q \rightarrow p_x$ and $p_q \rightarrow -x$ to arrive at the above Hamiltonian \eqref{HH1}. The above Hamiltonian \eqref{HH1} is identical to the one \eqref{HD1} obtained following Dirac constraint analysis. It is therefore understood that removal of the total derivative terms from the action under integration by parts, after introducing the auxiliary variable (as in HF), is as good as keeping the total derivative terms intact in Dirac's formalism.

\subsubsection{Modified Horowitz' Formalism (MHF):}

Here again we seek the phase-space structure of action \eqref{Anm1}, but only after controlling the total derivative terms. Let us therefore integrate action \eqref{Anm2} by parts. The total derivative terms $\Sigma_R$ and $\Sigma_{R_1}^2$ are eliminated in the process, and one ends up with action \eqref{AMH1}. Now substituting the auxiliary variable following Horowitz' prescription \cite{3H}

\be Q = {\partial A_{11}\over \partial \ddot z} = {18\beta\over N^3\sqrt z}\left({\ddot z} - {\dot N\over N}\dot z\right)\ee
into the action \eqref{AMH1} and integrating it by parts, the rest of the total derivative terms, viz. $Q\dot z = \Sigma_{R_2}^2$ is eliminated, and one obtains

\be\begin{split}\label{q}
A_{11} &= \int\Big[-\dot {\mathcal Q}\dot z-\frac{3f'\dot\phi\dot{z} \sqrt z}{N} - \frac{3 f{\dot z}^2}{2 N\sqrt z}+ 6 kN f \sqrt z -
\frac{N^3\sqrt{z}}{36\beta}{\mathcal Q}^2-\frac{\dot N\dot z\mathcal{Q}}{N}+ \frac{18\beta k\dot z^2}{Nz^{\frac{3}{2}}}+\frac{36\beta Nk^2}{\sqrt{z}} \\&
\hspace{3.4 in}+z^{\frac{3}{2}}\left(\frac{1}{2N}\dot\phi^2-VN\right)\Big]dt.\end{split}\ee
We use the same symbol $(Q)$ for the auxiliary variables, although they are different in different formalisms. Proceeding as before (e.g. using $q = NQ$ etc.), one finally ends up with the following Hamiltonian (see \cite{5kMH})

\be\label{MH1}\begin{split} H_{11}& = N\mathcal{H}_{11}\\& = N\left[x P_z + \frac{ \sqrt z {P_x}^2}{36\beta} +
\frac{P_{\phi}^2}{2z^{\frac{3}{2}}}+\frac{3x f'(\phi)P_{\phi}}{z} + 3f\sqrt z \Big(\frac{x^2}{2 z} - 2k \Big)
- \frac{18 k \beta}{\sqrt z} \Big(\frac{x^2}{z} + 2k\Big)+\frac{9f'^2x^2}{2\sqrt z}+Vz^{\frac{3}{2}}\right].\end{split}\ee
Thus, we obtain identical phase-space structure \eqref{HD11} obtained following Dirac's constrained analysis, and no inconsistency has been revealed at this level.

\subsection{Non-minimal coupling appearing with higher order term:}

Following Horowitz' Formalism (HF) one arrives at (see appendix B)

\be\begin{split}\label{HH2}
H_2 & = N\mathcal{H}_{2} = N\left[x p_z + {\sqrt z\over 36 f(\phi)} p_x^2 -\left({\alpha z\over 6f(\phi)} + 2k\right)p_x  +  {p_\phi^2\over 2z^{3\over 2}} + \left(V + {\alpha^2\over 4 f(\phi)}\right)z^{3\over 2}\right].\end{split}\ee
Thus we arrive at the same phase-space structure \eqref{D5}, obtained following Dirac's constrained analysis carried out in sub-subsection 2.2.1. Clearly again we notice that removal of the total derivative terms appearing in the action in HF gives exactly the same phase-space structure \eqref{D5} obtained following Dirac's algorithm without controlling these terms.\\

\noindent
On the contrary, following Modified Horowitz' Formalism (MHF), one obtains (see appendix B)

\be\label{HM2}\begin{split}& H_{22} = N \mathcal{H}_{22} = \\&N \left[x P_z + {\sqrt z P_x^2\over 36 f}  +
{P_\phi^2\over 2 z^{3\over 2}} + {36 k x f' P_\phi\over z^2} + \left({3\alpha\over 2 \sqrt z} -
{18k f \over z^{3\over 2}} +{648 k^2 f'^2 \over z^{5\over 2}} \right)x^2 - 6\alpha k \sqrt z -
{36k^2 f\over \sqrt z} + V(\phi) z^{3\over 2}\right], \end{split}\ee
which is identical to the one \eqref{H22} obtained following Dirac's constrained analysis followed in sub-subsection 2.2.2. In both the cases total derivative terms match with the surface terms obtained under variational principle. Everything seems to be consistent here too.

\subsection{Modified Einstein-Gauss-Bonnet-Dilatonic coupled action.}

\subsubsection{Horowitz' Formalism (HF):}

As before, to cast the Hamiltonian for the action \eqref{Agb} under consideration in the background of Robertson-Walker metric \eqref{RW}, following Horowitz' formalism, we find the auxiliary variable in view of the action \eqref{A12} as,

\be\label{QH} Q={\partial L\over \partial\ddot z} =\frac{3\alpha\sqrt z}{N}+ \frac{18\beta}{N\sqrt{z}}\Big(\frac{\ddot z}{N^2}-\frac{\dot z\dot N}{N^3}+2k\Big)+\frac{3\gamma(\phi)}{N\sqrt z}\Big(\frac{\dot z^2}{N^2 z} + 4 k\Big)\ee
and substitute it judiciously back into the action \eqref{A12} which under integration by parts yields

\be\begin{split}\label{15}
A_3 &= \int\Bigg[-\dot Q\dot z-\frac{N^3\sqrt{z}Q^2}{36\beta}-\frac{N\alpha^2 z^{\frac{3}{2}}}{4\beta}-\frac{\gamma(\phi)^2 \dot z^4}{4N^3\beta z^{\frac{5}{2}}} -\frac{4N\gamma(\phi)^2 k^2}{\beta \sqrt{z}}+\frac{N^2\alpha Qz}{6\beta}+2kQN^2\\&+\frac{\gamma(\phi)\dot z^2 Q}{6\beta z}+\frac{2kN^2\gamma(\phi)Q}{3\beta}-\frac{\alpha \gamma(\phi)\dot z^2 }{2\beta N\sqrt{z}}-\frac{Q\dot N\dot z}{N}-\frac{2k\alpha N\gamma(\phi)\sqrt {z}}{\beta}\\&-\frac{12k\gamma(\phi)\dot z^2}{Nz^{\frac{3}{2}}}-\frac{2k\gamma(\phi)^2\dot z^2}{N\beta z^{\frac{3}{2}}}-\frac{24k^2N\gamma(\phi)}{\sqrt z}-\frac{3\gamma(\phi)\dot z^4}{2N^3z^{\frac{5}{2}}}+\frac{z^{\frac{3}{2}}\dot\phi^2}{2N}-VNz^{\frac{3}{2}} \Bigg]dt,\end{split} \ee
where, the total derivative term $Q\dot z$ vanishes due to the boundary condition. The Hamiltonian has finally been furnished as \cite{5jMH},

\be\begin{split}\label{H3} H_{3}&=N{\mathcal{H}_{3}} = N\Bigg[x p_z+\frac{\sqrt{z}p_x^2}{36\beta}-
\left(\frac{\alpha z }{6\beta}+\frac{\gamma x^2 }{6\beta z}+\frac{2k\gamma}{3\beta}+2k\right)p_x+
\frac{p_{\phi}^2}{2z^{\frac{3}{2}}}+\left(\frac{\gamma^2}{4\beta z^{\frac{5}{2}}} +
\frac{3\gamma}{2z^{\frac{5}{2}}}\right) x^4\\& \hspace{0.7 in}+\frac{\alpha^2 z^{\frac{3}{2}}}{4\beta}+
\left(\frac{\alpha \gamma}{2\beta \sqrt{z}}+\frac{12k\gamma}{z^{\frac{3}{2}}} +\frac{2k\gamma^2}
{\beta z^{\frac{3}{2}}}\right)x^2+\frac{2\alpha k\gamma\sqrt{z}}{\beta}+\frac{24 k^2\gamma}
{\sqrt z}+\frac{4k^2\gamma^2}{\beta \sqrt{z}} +Vz^{\frac{3}{2}}\Bigg].\end{split}\ee
which is identical to the one \eqref{Hgbd} obtained following Dirac's constraint analysis without taking care of total derivative terms stored in the action.\\

\noindent
Mathematical inconsistency of this Hamiltonian due to the presence of $\gamma(\phi)$ term instead of $\gamma'(\phi)$ term has already been discussed. Here, we note yet another inconsistency, which remained in disguise while following Dirac's formalism. The Gauss-Bonnet-dilatonic part of the total derivative term appearing from $Q\dot z$, viz. ${3\gamma \sqrt z \dot z\over N^3}\left({\dot z^2\over z^2} + 4N^2{k\over z}\right)$, is different from the one $\Sigma_\mathcal{G} = {\gamma \sqrt z \dot z\over N^3}\left({\dot z^2\over z^2} + 12 N^2{k\over z}\right)$, obtained in view of variational principle. This means, in the process of canonical formulation, an additional term, viz. ${2\gamma\dot z^3\over N^3 z^{3\over 2}}$ has been thrown out of the action, which is in direct contradiction to the variational principle. This is clearly a mathematical inconsistency, which as mentioned, remains in disguise in Dirac's constrained analysis made in sub-subsection 2.3.1.

\subsubsection{Modified Horowtz' Formalism (MHF):}

As already understood, in MHF, following integration by parts, one starts with action \eqref{A20} and the auxiliary variable

\be\label{QM} \mathcal Q=\frac{18 \beta}{N^3\sqrt z}\left({\ddot z}-\frac{\dot N \dot z}{ N}\right)\ee
is introduced at this stage straight into the action \eqref{A20} and integrated by parts again, so that the rest of the total derivative terms, viz. $\Sigma_{R^2_2}$ is eliminated, to end up with,

\be\begin{split}\label{A21}
A_{33} = \int\bigg[-\dot {\mathcal Q}\dot z-\frac{N^3\sqrt{z}}{36\beta}{\mathcal Q}^2-\frac{\dot z\dot N \mathcal Q}{N} &-\frac{3\alpha\dot z^2}{2N\sqrt z}+6\alpha kN\sqrt{z} + \frac{18\beta k\dot z^2}{Nz^{\frac{3}{2}}} +\frac{36\beta Nk^2}{\sqrt{z}} -\frac{\gamma'(\phi)\dot\phi{\dot z}^3}{N^3z^{\frac{3}{2}}}\\& -\frac{12k\gamma'(\phi)\dot\phi\dot z}{N\sqrt{z}}+z^{\frac{3}{2}}\left(\frac{1}{2N}\dot\phi^2-VN\right)\bigg]dt.\end{split}\ee
The Hamiltonian has finally been formulated as \cite{5jMH},

\be\begin{split}\label{H33}
H_{33}  = N{\mathcal H_{33}}= N\bigg[x P_z &+ \frac{ \sqrt z {P_x}^2}{36\beta} +\frac{P_{\phi}^2}{2z^{\frac{3}{2}}}+\left(\frac{x^3}{z^3}+\frac{12kx}{z^2}\right)\gamma'P_{\phi}+ 3\alpha\Big(\frac{x^2}{2 \sqrt z} - 2k \sqrt z \Big)\\& - \frac{18 k \beta}{\sqrt z} \Big(\frac{x^2}{z} + 2k\Big)+\left(\frac{x^6}{2z^{\frac{9}{2}}}+\frac{12kx^4}{z^{\frac{7}{2}}}+ \frac{72k^2x^2}{z^{\frac{5}{2}}}\right)\gamma'^2+Vz^{\frac{3}{2}} \bigg],
\end{split}\ee
which is identical to \eqref{Hgbm}. Here the total derivative terms match with the surface terms derived in view of variational principle, and everything is consistent. This fact reviles that the total derivative terms indeed are a big issue for canonical formulation of higher-order theories.

\section{Concluding Remarks}

It has been demonstrated that all the constrained Hamiltonian formulations of higher derivative theories are canonically equivalent once they are constructed by introducing different variables absorbing higher derivatives of the original coordinates \cite{Buch}. Further, canonical relationships between various Hamiltonian formulations of $f(R)$ theory of gravity (in Einstein's frame, Jordan's frame or following Ostrogradski's technique) have been found to exist \cite{deurel}. The present manuscript does not differ in this respect. However, equivalence at the quantum level requires canonical transformation in quantum mechanics \cite{Arlen}. Now, since at the classical level canonical transformations are highly non-linear, so the different quantum descriptions so obtained are not likely to be equivalent. In this connection we explore an important issue that has not been dealt with as yet. This is the way to pick one Hamiltonian (out of several canonically equivalent ones) for quantization. The issue is important because, canonically equivalent Hamiltonians produce different quantum descriptions which although are related through unitary transformation, such transformation is not one-to-one for non-minimally coupled higher order theories. \\

In the minimal case studied earlier \cite{5jMH}, canonical quantization of the two canonically related Hamiltonians obtained following two techniques (HF and MHF) produce two different effective potentials. While, the extremum of the effective potential obtained following MHF lead to $a = a_0 e^{2\sqrt{2 V\over 3\alpha}t}$, revealing an interesting fact that inflation is a generic feature of higher order theory of gravity, the extremum of the effective potential obtained form HF only yields $V(\phi) = -{\alpha^2\over 4\beta} = -V_0$, which is a constant. It's true that the two are related under unitary transformation, but how can one make such transformation unless he follows the route towards MHF? Additionally, if one follows MHF, what is the reason to make a unitary transformation to obtain quantum dynamics followed by HF? This fact led us to choose the Hamiltonian obtained following MHF, (or equivalently following Dirac's constrained analysis, taking care of the total derivative terms present in the action a-priory) for quantization. Note that the identification of the degree of a classical action requires integration by parts. For example, general theory of relativity appears to be a higher-order theory. Only under integration by parts a total derivative term is eliminated, leading it to a standard second order theory. Likewise, integration by parts is required to perform for checking the degree of a theory. This has been explicitly exemplified earlier \cite{br2}. The question is why should one integrate the action by parts only to reduce a theory from higher to lower order and keep total derivative terms of the lower order intact?\\

To make our arguments of picking up one of the two Hamiltonians (obtained following MHF or equivalently following Dirac's algorithm taking care of total derivative terms a-priori) to be suitable for quantization, more convincing, and to exhibit the difference between the two predominantly, we have taken up three cases of non-minimally coupled higher order theories of gravity, and constructed phase-space structures without and with taking care of the total derivative terms a-priori. We list below the findings.
\begin{enumerate}
\item Dirac's constrained analysis without taking care of total derivative terms present in the action and HF produce the same Hamiltonian. On the contrary, Dirac's constrained analysis taking care of total derivative terms present in the action and MHF produce identical Hamiltonian.
\item In all the three cases the two Hamiltonians are found to be canonically equivalent. This once again prove that total derivative term appearing in an action does not change the equation of motion, rather it just changes the canonical momenta in a way that only amounts to a canonical transformation.
\item The difference is predominant while attempting to quantize. While the coupling parameter has its shear presence in the first Hamiltonian, in the second, its presence requires operator ordering. Thus different forms of the coupling parameter lead to different quantum descriptions.
\item In the Gauss-Bonnet-dilatonic coupled case also, the Hamiltonians are canonically related and as such, classical field equations remain unchanged. Nevertheless, a mathematical inconsistency is exhibited in the Hamiltonian derived in the first case. Gauss-Bonnet term is topologically invariant in four-dimensions, and so it does not contribute to the field equations. This is why, a dilatonic coupling $\gamma(\phi)$ is required. Now if $\gamma$ turns out to be a constant, it must not contribute in the quantum domain also, as is clearly seen from \eqref{Hgbm}. However, in the first case \eqref{Hgbm} Gauss-Bonnet term contributes, even if the coupling parameter is constant.
\item In the HF, to construct phase-space structure for MEGBD, it has been found that an additional term (${2\gamma\dot z^3\over N^3z^{3\over 2}}$) is thrown away as a boundary term, which contradicts the variational principle. This is the second mathematical inconsistency.
\end{enumerate}

\noindent
In view of the above results, we conclude that, either MHF or Dirac's constrained analysis only after taking care of total derivative terms yields the correct quantum description of higher order theories. A final comment. Also we have proved that once the action is made free from total derivative terms, Dirac formalism is identical to MHF. However, there is absolutely no scope to remove the total derivative term $\Sigma_{R_2^2}$ from the action, once Dirac's formalism is initiated. While MHF, takes care of all the total derivative terms which are at par with variational principle, in two steps. Thus, it's simply a matter of taste for choosing a formalism between the two.

\appendix

\section{Detailed calculation of subsection 2.1 }

In the appendix, we explicitly explore the viability of the Hamiltonian \eqref{HD1} so obtained, in classical, quantum and semiclassical domains.\\

\noindent
\textbf{Classical domain:}\\

\noindent
Firstly, one can find the Hamilton's equations as,

\be\label{HE1}\begin{split}& \dot z = x;\;\;\dot p_z = -{p_x^2\over 72\beta\sqrt z} +{f p_x\over 6\beta}
+ {3p_\phi^2\over 4 z^{5\over 2}} -{3f^2\sqrt z\over 8\beta}-{3\over 2}V\sqrt z;\\&
\dot x = {\sqrt z\over 18\beta}p_x - {fz\over 6\beta} - 2k;\;\;\dot p_x = - p_z;\;\;\;\;\;\dot\phi =
{p_\phi\over z^{3\over 2}};\;\;\dot p_\phi = {z f'\over 6\beta}p_x - {ff' z^{3\over 2}\over 2\beta} - V' z^{3\over 2},
\end{split}\ee
which upon  little algebraic manipulation lead to

\be\label{m1} p_\phi = z^{3\over 2}\dot \phi, \;\;\;p_x = {18\beta\over \sqrt z}(\ddot z + 2k) + 3f\sqrt z,\;\;\; p_z
= -{18\beta\over \sqrt z}\dddot z + {9\beta\dot z\over z^{3\over 2}}(\ddot z + 2k) - {3f\dot z\over 2\sqrt z}
- 3\sqrt z f'\dot\phi.\ee
Upon substituting the above expressions for momenta and replacing $z$ by $a^2$ in the Hamiltonian (\ref{HD1}),
the $(^0_0)$ equation of Einstein \eqref{01} is obtained. One can also find the scalar field equation \eqref{01}
as well, in view of the last pair of Hamilton's equations \eqref{HE1}. This proves that the Hamiltonian \eqref{HD1}
so obtained, is correct from classical point of view. Further, the action \eqref{Anm2} may also be expressed in
canonical (ADM) form as

\be\label{canact1} A_{1} = \int(\dot z p_z + \dot x p_x + \dot\phi p_{\phi} - N\mathcal{H}_{1D})~ dt d^3 x
= \int(\dot h_{ij} p^{ij} + \dot K_{ij}\pi^{ij} + \dot\phi p_{\phi} - N\mathcal{H}_{1D})~ dt d^3 x, \ee
where, $p_{ij}$ and $\pi^{ij}$ are the momenta canonically conjugate to $h_{ij}$ and $K_{ij}$ respectively.
It is now required to check the viability of the Hamiltonian \eqref{HD1} in the quantum domain, which has not
been tested earlier. This we pose underneath.\\

\noindent
\textbf{Quantum domain:}\\

\noindent
The hermiticity of the effective Hamiltonian appearing in \eqref{Ds} is ensured for $n = -1$, which allows one to write the continuity equation as,

\be \frac{\partial\rho}{\partial\sigma}+\nabla.\mathrm{\mathbf{J}}=0,\ee
where, $\rho=\Psi^{*}\Psi$ and $\mathrm{\mathbf{J}} = (J_x, J_{\phi}, 0)$ are the probability density and the current
density respectively, with, $J_x = \frac{i\hbar}{54\beta x}(\Psi^{*}_{,x}\Psi-\Psi^{*}\Psi_{,x})-
\left(\frac{2k}{3\sigma^{\frac{1}{3}}}+\frac{\alpha\sigma^{\frac{1}{3}} }{18\beta}\right)\frac{\Psi^{*}\Psi}{x}$
and $J_{\phi}=\frac{i\hbar}{3x\sigma^{\frac{4}{3}}}(\Psi^{*}_{,\phi}\Psi-\Psi^{*}\Psi_{,\phi})$.
In the process, probabilistic interpretation becomes straight-forward for higher order theory of gravity under
consideration.\\

\noindent
\textbf{Semiclassical approximation:}\\

\noindent
To further test the authenticity of the Hamiltonian \eqref{HD1} in the quantum domain, we perform
semiclassical approximation. To avoid unnecessary complication, we express wave-equation \eqref{Dq} in the following
form ($k = 0,~ n = -1$)

\be\label{Dsc}
-\frac{\hbar^2 \sqrt z}{36\beta x}\left(\frac{\partial^2}{\partial x^2} - \frac{1}{x}\frac{\partial}{\partial x}\right)\Psi
+ i\hbar\left({fz\over 6 \beta x}{\partial \Psi\over \partial x} - {\partial\Psi\over \partial z}\right)
- {\hbar^2\over 2 x z^{3\over 2}}\left({\partial^2\Psi\over \partial\phi^2}\right)+ \mathcal{V}\Psi = 0,\ee
where, $\mathcal{V} = {z^{3\over 2}V\over x} + {f^2 z^{3\over 2}\over 4\beta x} - {i\hbar f z\over 12\beta x^2}$.
The above equation may be treated as time independent Schr\"{o}dinger equation with three variables $x$, $z$ and $\phi$.
Hence as usual, let us seek the solution of the wave-function (\ref{Dsc}) as,

\be\label{Psi1} \Psi=\Psi_0 e^{\frac{i}{\hbar}S(x,z,\phi)}\ee
and expand $S$ in power series of $\hbar$ as,

\be\label{Sx1} S = S_0(x, z,\phi) + \hbar S_1(x, z, \phi) + \hbar^2S_2(x, z, \phi) + .... .\ee
One can then find,

\bes\label{psi2}
\Psi_{,x} = {i\over \hbar}[S_{0,x}+\hbar S_{1,x}+\hbar^2 S_{2,x}+\mathcal{O}(\hbar)]\Psi;~~~~~\Psi_{,xx}
= {i\over \hbar}[S_{0,xx} + \hbar S_{1,xx} + \hbar^2 S_{2,xx}+ \mathcal{O}(\hbar)]\Psi\\&\hspace{0.34 in}
-{1\over \hbar ^2}[S_{0,x}^2 + \hbar^2 S_{1,x}^2 + \hbar^4 S_{2,x}^2 + 2\hbar S_{0,x} S_{1,x} + 2\hbar^2 S_{0,x} S_{2,x}
+ 2\hbar^3 S_{1,x} S_{2,x}+\mathcal{O}(\hbar)]\Psi;\\&\Psi_{,z} = {i\over \hbar}[S_{0,z}+\hbar S_{1,z}+\hbar^2 S_{2,z}
+\mathcal{O}(\hbar)]\Psi;~~~~~\Psi_{,\phi\phi} = {i\over \hbar}[S_{0,\phi\phi} + \hbar S_{1,\phi\phi}
+ \hbar^2 S_{2,\phi\phi}+\mathcal{O}(\hbar)]\Psi\\&\hspace{0.34 in} -{1\over \hbar ^2}[S_{0,\phi}^2
+ \hbar^2 S_{1,\phi}^2 + \hbar^4 S_{2,\phi}^2 + 2\hbar S_{0,\phi} S_{1,\phi} + 2\hbar^2 S_{0,\phi} S_{2,\phi}
+ 2\hbar^3 S_{1,\phi} S_{2,\phi}+\mathcal{O}(\hbar)]\Psi,\end{split}\ee
etc., where ``comma" in the suffix stands for derivative. Now, inserting $\Psi, \Psi_{,x}, \Psi_{,xx}, \Psi_{,z},
\Psi_{,\phi\phi}$ etc. in view of \eqref{Psi1}, and \eqref{psi2} in equation \eqref{Dsc} and equating the coefficients
of different powers of $\hbar$ to zero, the following set of equations (upto second order) are obtained.

\begin{subequations}\begin{align}
&\label{S0}\frac{\sqrt z}{36\beta x}S_{0,x}^2+\frac{S_{0,\phi}^2}{2xz^{\frac{3}{2}}}+S_{0,z}-\frac{f z}{6\beta x}S_{0,x}
+ \left({f^2\over 4 \beta} +V\right) {z^{3\over 2}\over x}= 0\\
&\label{S1}i\left[\frac{\sqrt z S_{0,xx}}{36\beta x} - \frac{\sqrt z S_{0,x}}{36\beta x^2}
+ \frac{S_{0,\phi\phi}}{2xz^{\frac{3}{2}}} + {f z\over \beta x^2} \right]-S_{1,z}
-\frac{\sqrt z S_{0,x}S_{1,x}}{18\beta x}-\frac{S_{0,\phi}S_{1,\phi}}{xz^{\frac{3}{2}}}+\frac{f z}{6\beta x}S_{1,x}=0\\
&i\left[\frac{\sqrt z S_{1,xx}}{36\beta x}-\frac{\sqrt z S_{1,x}}{36\beta x^2}
+\frac{S_{1,\phi\phi}}{2xz^{\frac{3}{2}}}\right]-S_{2,z}+\frac{\sqrt zS_{0,x}S_{2,x}}{18\beta x}
-\frac{S_{0,\phi}S_{2,\phi}}{xz^{\frac{3}{2}}}-{\sqrt z S_{1,x}^2\over 36\beta x}-{S_{1,\phi}^2\over 2x z^{3\over 2}}
+\frac{f z S_{2,x}}{6\beta x}=0
\end{align}\end{subequations}
which are to be solved successively to find $S_0(x, z,\phi)$, $S_1(x, z,\phi)$ and $S_2(x, z,\phi)$ and so on.\\

\noindent
\textbf{First consistency check:}\\

\noindent
Identifying $S_{0,x}$ with $p_x$, $S_{0,z}$ with $p_z$ and $S_{0,\phi}$ with $p_\phi$ respectively, the Hamilton constraint equation \eqref{HD1} is retrieved. Further, using the definitions of momenta \eqref{m1}, equation (\ref{S0}) is expressed as,

\be -6f{\dot a^2\over a^2} - 6{\dot a\over a} f'\dot \phi - {36\beta\over a^2} \left[2\dot a\dddot a - \dot a^2
+ 2{\dot a^2\ddot a\over a} - 3{\dot a^2\over a^2}\right] + \left({1\over 2}\dot\phi^2 + V\right) = 0.\ee
In the process $(^0_0)$ component of Einstein's equation \eqref{00} (for $N= 1, k =0)$ has been retrieved and hence
the Hamiltonian \eqref{HD1} is successfully through with the first consistency check.\\

\noindent
\textbf{Second consistency check:}\\

\noindent
In view of the classical solutions \eqref{sol1} one can now compute the momenta \eqref{m1} and their integrals as

\be\label{momall}\begin{split}&
p_x = \Big({3f_0\over\sqrt{2\Lambda}} + 36\sqrt{2}\beta \Lambda^{3\over 2}\Big)\sqrt x
+ {3 f_1 x\over 2\Lambda a_0\phi_0} - {\sqrt{\Lambda} a_0^2\phi_0^2\over 2\sqrt {2x}};\\& p_z
= -3\Lambda\Big(f_0+24\beta \Lambda^2\Big)\sqrt z - {6\Lambda f_1 z\over a_0\phi_0}
- {\Lambda a_0^2\phi_0^2\over 4\sqrt z};\hspace{0.3 in}p_\phi = -{\Lambda a_0^3\phi_0^3\over \phi^2}\\&
\int p_x dx = \sqrt{2\over \Lambda}\Big(f_0+24\beta \Lambda^2\Big)x^{3\over 2}
- {3 f_1 x^2\over 4\Lambda a_0\phi_0}-{\sqrt{\Lambda \over 2}a_0^2\phi_0^2}\sqrt x
= \Big[4\Big(f_0 + 24\beta \Lambda^2\Big) z^{3\over 2} + {3f_1  z^2\over a_0\phi_0}
-  a_0^2\phi_0^2\sqrt z\Big]\Lambda\\&
\int p_z dz = -2\Lambda\Big(f_0 - 24\beta \Lambda^2\Big) z^{3\over 2} - {3f_1 \Lambda z^2\over a_0\phi_0}
- {\Lambda a_0^2\phi_0^2\sqrt z\over 2};~~~
\int p_\phi d\phi = \frac{\Lambda a_0^3\phi_0^3}{\phi}= \Lambda a_0^2\phi_0^2\sqrt z
\end{split}\ee
Thus $S_{0}$, which when expressed in terms of the integrals of momenta yields

\be \label{HJ1} S_{0} = \int p_x dx + \int p_z dz + \int p_\phi d\phi = 2 f_0\Lambda z^{3\over 2}
+ 48\beta \Lambda^3 z^{3\over 2} - {\Lambda a_0^2\phi_0^2\over 2}\sqrt z.\ee
One can also compute the zeroth order on-shell action \eqref{Anm2} in view of the classical solutions
\eqref{sol1} as

\be \label{HJ2} A_{1\mathrm{Cl}} = \int \left[3 f_0 \Lambda \sqrt z + 72\beta\Lambda^3\sqrt z -
{\lambda a_0^2\phi_0^2\over 4\sqrt z}\right] dz = 2 f_0\Lambda z^{3\over 2} + 48\beta \Lambda^3 z^{3\over 2}
- {\Lambda a_0^2\phi_0^2\over 2}\sqrt z.\ee
Since the classical on-shell action is identical with the Hamilton-Jacobi function, so the Hamiltonian \eqref{HD1}
is also successfully through to the second consistency check. At this end, one can express the wave function as,

\be\label{psi0} \Psi_{1} = \psi_{01} e^{{i\over \hbar}\Lambda\left[\big(48\beta \Lambda^2 + 2 f_0\big)z^{3\over 2}
- {a_0^2\phi_0^2 \sqrt z \over 2}\right]}.\ee
It is now in principle possible to solve equation \eqref{S1} in the form $S_1 = i G_1(z)$ on the solutions \eqref{sol1}, and therefore the wavefunction may be expressed upto the first order approximation as presented in \eqref{psi1}. Thus, first-order approximation only modifies the pre-factor of the wavefunction, keeping the oscillatory behavior of the wave function unaltered. The oscillatory behaviour of the wavefunction indicates that the region is classically allowed and the wavefunction is strongly peaked about a set of exponential solutions \eqref{sol1} to the classical field equations \eqref{00}. This establishes the correspondence between the quantum equation and the classical equations, resulting in a viable quantum theory. So altogether, for the system \eqref{Anm1}, the Hamiltonian \eqref{HD1}
obtained following Dirac's constraint analysis, keeping the total derivative terms intact, is particularly well-behaved. \\

\section{Detailed Calculation of subsection 3.2}
\subsection{Horowitz' Formalism (HF):}

We have proved that Dirac formalism in the presence of total derivative terms yields identical Hamiltonian as obtainable from HF. Here, we first explicitly derive the Hamiltonian in connection with the action \eqref{A2}, following HF, and then prove its viability, which was referred to in sub-subsection-2.2.1. As already understood, in HF, it is required to find the auxiliary variable right from the action \eqref{A2} as,

\be Q = {\partial A\over \partial \ddot z} = {3\alpha\sqrt z\over N} + {18 f\over N\sqrt z}\left({\ddot z\over N^2} - {\dot N\dot z\over N^3} + {2k}\right),\ee
and insert it judiciously in the action \eqref{A2} as,

\be A_2 = \int\left[Q\ddot z - {N^3\sqrt z\over 36 f}Q^2 + 2kN^2 Q + {\alpha N^2 Q z\over 6 f} - {\dot N\dot zQ\over N} -{N\alpha^2 z^{3\over 2}\over 4 f} +\Big({1\over 2 N}\dot \phi^2  - N V\Big) z^{3\over 2}  \right]dt.\ee
Now, integrating it by parts, one obtains

\be A_2 = \int\left[-\dot Q\dot z - {N^3\sqrt z\over 36 f}Q^2 + 2kN^2 Q + {\alpha N^2 Q z\over 6 f} - {\dot N\dot zQ\over N} -{N\alpha^2 z^{3\over 2}\over 4 f} +\Big({1\over 2 N}\dot \phi^2  - N V\Big) z^{3\over 2}  \right]dt\ee
where, the total derivative term,

\be \label{a1} Q\dot z = {3\alpha\sqrt z\dot z\over N} + {36 f\sqrt z \dot z\over N^3}\left({\ddot z\over 2z} + + {N^2k\over z}- {\dot N\dot z\over 2 N z} \right)\ee
exactly matches with the total derivative terms that appear from the variation of action \eqref{A2}. The canonical momenta are

\be\label{a2} p_z =-\dot Q-{\dot N Q\over N};\;\;\; p_\phi = {{z^{\frac{3}{2}}\dot\phi}\over N};\;\;\; p_Q =
  -\dot z;\;\;\;p_N=-\frac{\dot z Q}{N}.
\ee
Although, a constraint in the form $Qp_Q - Np_N = 0$ is apparent, still one can readily find the Hamiltonian in phase space variables without requiring constraint analysis. To avoid the said constraint, it is better to choose a more fundamental variable, $q = NQ$ as mentioned in sub-subsection 3.1.1. In view of this variable, the action and the canonical momenta read as,

\bes A_2 = \int\left[-{\dot q\dot z\over N} - {N\sqrt z\over 36 f}q^2 + 2kN q + {\alpha N Q z\over 6 f} -{N\alpha^2 z^{3\over 2}\over 4 f} +\Big({1\over 2 N}\dot \phi^2  - N V\Big) z^{3\over 2} \right]dt,\\&
p_q = -{\dot z\over N},~~~~~p_z = -{\dot q\over N},~~~~~p_\phi = {\dot \phi\over N}z^{3\over 2},\end{split}\ee
and the constraint disappears. The Hamiltonian may now readily be expressed as

\be\label{a3}
 H_2 =N\left[-p_q p_z + \frac{\sqrt{z}q^2}{36 f}-\left(2k + {\alpha z\over 6f}\right)q +\frac{ p^2_{\phi}}{2z^{\frac{3}{2}}}+\left( V + {\alpha^2\over 4f}\right) z^{\frac{3}{2}}\right],\ee
and diffeomorphic invariance is thereby established, where $N$ acts as a Lagrange multiplier only. As before, it is now required to express the Hamiltonian in terms of basic variables ($z$ and $x =\frac{\dot z}{N}$), instead of ($z$ and $q$). Since, $p_q =-{\dot z\over N} = -x$ and $q = p_x$, one therefore is required to make the following canonical transformation under the replacements, $p_q$ by $-x$ and $q$ by $p_x$ in the Hamiltonian (\ref{a3}), which therefore finally results in,

\be\begin{split}\label{a4}
H_2 & = N\mathcal{H}_{2} = N\left[x p_z + {\sqrt z\over 36 f(\phi)} p_x^2 -\left({\alpha z\over 6f(\phi)} + 2k\right)p_x  +  {p_\phi^2\over 2z^{3\over 2}} + \left(V + {\alpha^2\over 4 f(\phi)}\right)z^{3\over 2}\right],\end{split}\ee
It is noticeable yet again that the presence of the coupling $f(\phi)$ in the action \eqref{Anm3} has only its shear presence in the Hamiltonian \eqref{a4}, and doesn't affect its form from the one obtained with constant coupling, (i.e. it may be obtained simply by replacing $\beta$ by $f(\phi)$ in equation (35) of \cite{5jMH}). One can again trivially check that in view of the Hamilton's equations obtained from the above Hamiltonian \eqref{a4} the field equations \eqref{FE} are recovered. The action \eqref{A2} may also be expressed in canonical form in the following manner. In view of the Hamilton's equations,

\be\label{a5}
 p_x=3{\alpha\sqrt z}+\frac{18 f(\phi)}{N\sqrt{z}}\left({\dot x} + 2k N\right),\;\;\;
\dot p_x=-\frac{\partial \mathcal H_2}{\partial x}=-Np_z.
\ee
Further, in view of the Hamiltonian (\ref{a4}) one finds,

\be \begin{split}&\left(\dot z p_z + \dot x p_x +\dot\phi p_{\phi}- N\mathcal{H}_{2}\right)=\dot x p_x
+\dot\phi p_{\phi}-\left[\frac{N\sqrt z}{36f(\phi)}p_x^2-2kNp_x-\frac{N\alpha z p_x}{6f(\phi)}
+\frac{N\alpha^2z^{\frac{3}{2}}}{4f(\phi)}+\frac{p_{\phi}^2}{2z^{\frac{3}{2}}}+Vz^{\frac{3}{2}}\right]\\&
=3{\alpha\sqrt z\dot x}+\frac{18f(\phi)\dot x}{N\sqrt{z}}\left({\dot x} + 2k N\right)
-\left[\frac{9f(\phi)}{N\sqrt z}\left({\dot x} + 2k N\right)^2-6k\alpha N\sqrt z-\frac{36f(\phi) k}{\sqrt z}
\left({\dot x} + 2k N\right)\right]\\& \hspace{0.6 in}+z^{\frac{3}{2}}\left(\frac{\dot\phi^2}{2N}-VN\right)
=\Bigg[3\alpha\sqrt z\left(\dot x+2kN\right)+\frac{9f(\phi)}{N\sqrt z}\left(\dot x+2kN\right)^2+z^{\frac{3}{2}}
\left(\frac{\dot\phi^2}{2N}-VN\right)\Bigg],\end{split}\ee
and therefore, the action (\ref{A2}) can now be expressed in the canonical form in terms of the basic variables as,

\be \label{a6}A_2 = \int\left( \dot z p_z + \dot x p_x+\dot\phi p_{\phi} - N\mathcal{H}_{2} \right)dt~ d^3 x\, \ \
= \int\left(\dot h_{ij} p^{ij} + \dot K_{ij}\pi^{ij} +\dot\phi p_{\phi}- N\mathcal{H}_{2}\right)dt~ d^3 x,\ee
where, $p^{ij}$ and $\pi^{ij}$ are momenta canonically conjugate to $h_{ij}$ and $K_{ij}$ respectively. Thus,
canonical formulation of the higher order theory of gravity \eqref{Anm3} under consideration has been performed
in Robertson-Walker minisuperspace background \eqref{RW}.\\

\noindent
\textbf{Canonical Quantization:}\\

\noindent
Since due to diffeomorphic invariance the Hamiltonian \eqref{a4} is constrained to vanish, so canonical quantization leads to

\be\label{a7}\begin{split}
\frac{i\hbar}{\sqrt z}\frac{\partial \Psi}{\partial z} = -\frac{\hbar^2}{36f(\phi) x}
\left(\frac{\partial^2}{\partial x^2}+ \frac{n}{x}\frac{\partial}{\partial x}\right)\Psi &
+i\hbar\left(\frac{2k}{\sqrt z}+\frac{\alpha \sqrt z}{6f(\phi)}\right)\frac{1}{2}\left(\frac{2}{x}
\frac{\partial\Psi}{\partial x}-\frac{\Psi}{x^2}\right)\\& \hspace{0.4 in}-\frac{\hbar^2}{2xz^2}
\frac{\partial^2\Psi}{\partial\phi^2}+{z\over x}\left(V(\phi)+\frac{\alpha^2}{4f(\phi)}\right)\Psi,\end{split}\ee
where Weyl symmetric operator ordering has been performed in the 1st. and the 3rd. terms appearing on right hand side, $n$ being the operator ordering index. Under a further change of variable, the above modified Wheeler-de-Witt equation, takes the look of Schr\"odinger equation, viz.,

\be\label{a8} i\hbar \frac{\partial \Psi}{\partial\sigma}=-\frac{\hbar^2}{54f(\phi)}\left(\frac{1}{x}
\frac{\partial^2}{\partial x^2} + \frac{n}{x^2}\frac{\partial}{\partial x}\right)\Psi-
\frac{\hbar^2}{3x\sigma^{\frac{4}{3}}}\frac{\partial^2\Psi}{\partial\phi^2}
+i\hbar\left(\frac{4k}{3\sigma^{\frac{1}{3}}}+\frac{\alpha\sigma^{\frac{1}{3}} }{9f(\phi)}\right)
\left(\frac{1}{x}\frac{\partial}{\partial x}-\frac{1}{2 x^2}\right) \Psi+V_e\Psi=\hat{H_e}\Psi\ee
where, the proper volume, $\sigma=z^{\frac{3}{2}}=a^3$ plays the role of internal time parameter. In the above, $\hat H_e$ is the effective Hamiltonian operator and $V_e={2\sigma^{\frac{2}{3}}\over 3x}\big(V+\frac{\alpha^2 }{4f(\phi)}\big)$ is the effective potential. The hermiticity of the effective Hamiltonian is ensured for $n = -1$, which enables one to write the continuity equation as,

\be   \frac{\partial\rho}{\partial\sigma}+\nabla.\mathrm{\mathbf{J}}=0,\ee
where, $\rho=\Psi^{*}\Psi$ and $\mathrm{\mathbf{J}} = (J_x, J_{\phi}, 0)$ are the probability density and the current
density respectively, with, $J_x = \frac{i\hbar}{54f(\phi) x}(\Psi^{*}_{,x}\Psi-\Psi^{*}\Psi_{,x})
-\left(\frac{2k}{3\sigma^{\frac{1}{3}}}+\frac{\alpha\sigma^{\frac{1}{3}} }{18f(\phi)}\right)\frac{\Psi^{*}\Psi}{x}$
and $J_{\phi}=\frac{i\hbar}{3x\sigma^{\frac{4}{3}}}(\Psi^{*}_{,\phi}\Psi-\Psi^{*}\Psi_{,\phi})$.
In the process, probabilistic interpretation becomes straight-forward for higher order theory of gravity under
consideration following HF.\\

\noindent
\textbf{Semiclassical approximation:}\\

\noindent
Since we have a set of classical solutions \eqref{sol} at hand for flat space, therefore to perform semiclassical
approximation, we take up the quantum equation \eqref{a7}, set $k = 0$ and express it as,
\be\label{a9}\begin{split}&
 -\frac{\hbar^2\sqrt z}{36 f x}\left(\frac{\partial^2}{\partial x^2} + \frac{n}{x}\frac{\partial}{\partial x}\right)\Psi
 +i\hbar\left[\frac{\alpha z}{6 f x}\frac{\partial\Psi}{\partial x}-\frac{\partial \Psi}{\partial z}\right]
 -\frac{\hbar^2}{2xz^{\frac{3}{2}}}\frac{\partial^2\Psi}{\partial \phi^2} +\mathcal V\Psi=0,\\&
\mathrm{where},\;\;\mathcal V = {z^{3\over 2}\over x}V + {\alpha^2 z^{3\over 2}\over 4 f x}
-{i\hbar \alpha z\over 12 f x^2} .\end{split}\ee
The above equation may be viewed as time independent Schr\"{o}dinger equation with three variables $x$, $z$ and $\phi$. Hence, as before, let us seek the solution of the wave-equation as in \eqref{Psi1}, and insert $\Psi, \Psi_{,x}, \Psi_{,xx}, \Psi_{,z},
\Psi_{,\phi\phi}$ etc. in view of \eqref{Psi1}, and \eqref{psi2} in equation \eqref{a9} and equate the coefficients
of different powers of $\hbar$ to zero, to obtain the following set of equations (upto second order).

\begin{subequations}\begin{align}
&\label{s0}\frac{\sqrt z}{36 f x}S_{0,x}^2+ \frac{S_{0,\phi}^2}{2xz^{\frac{3}{2}}} - \frac{\alpha z S_{0,x}}{6 f x}
+ S_{0,z} + {z^{3\over 2}\over x}V + {\alpha^2 z^{3\over 2}\over 4f x}= 0\\
&\label{q11}i\left[\frac{\sqrt z}{36 f x}S_{0,xx}+ \frac{n\sqrt z}{36 f x^2}S_{0,x}
+ \frac{S_{0,\phi\phi}}{2xz^{\frac{3}{2}}} + {\alpha z\over 12 f x^2}\right] -\frac{\sqrt zS_{0,x}S_{1,x}}{18 f x}
-\frac{S_{0,\phi}S_{1,\phi}}{xz^{\frac{3}{2}}}+\frac{\alpha z}{6 f x}S_{1,x}-S_{1,z}=0\\
&i\left[\frac{\sqrt z S_{1,xx}}{36 f x}+\frac{n\sqrt z S_{1,x}}{36 f x^2}+\frac{S_{1,\phi\phi}}{2xz^{\frac{3}{2}}}\right]
-{\sqrt z S_{1,x}^2\over 36 f x}-\frac{\sqrt zS_{0,x}S_{2,x}}{18 f x}+\frac{\alpha z S_{2,x}}{6 f x}
-\frac{S_{0,\phi}S_{2,\phi}}{xz^{\frac{3}{2}}} - {S_{1,\phi}^2\over 2x z^{3\over 2}}-S_{2,z}=0
\end{align}\end{subequations}
which are to be solved successively to find $S_0(x, z,\phi)$, $S_1(x, z,\phi)$ and $S_2(x, z,\phi)$ and so on. Now
identifying $S_{0,x}$ with $p_x$, $S_{0,z}$ with $p_z$ and $S_{0,\phi}$ with $p_\phi$, one can obtain the Hamilton
constraint equation \eqref{a4}. Further in view of the definition of canonical momenta $p_x$ \eqref{a5} and $p_{\phi}$ \eqref{a2} or (107) and $p_z = -\dot p_x$ \eqref{a5}, it is also possible to regain the time-time component of Einstein's equation
\eqref{FE} in flat space. So far so good, since everything is consistent and there is no problem as such. Now, in order
to compute $S_0(x, z, \phi)$ let us express it as,

\be\label{S} S_0=\int p_z dz+\int p_x dx+\int p_\phi d\phi\ee
apart from a constant of integration which may be absorbed in $\Psi_0$. In view of the classical solution \eqref{sol},
it is possible to interrelate all the variables. For example,
\be \label{rel} x = \dot z = 2\Lambda z,\;\; \phi = {a_0\phi_0\over \sqrt z},\;\;f =f_0 +
{V_1\over 144\Lambda^4a_0\phi_0}\sqrt z - {a_0^2\phi_0^2\over 288\Lambda^2 z},\;\;V = 6\alpha\Lambda^2
+ {V_1\sqrt z\over a_0\phi_0} + {\Lambda^2 a_0^2\phi_0^2\over 2z}.\ee
and so on. The integrals in the above expression \eqref{S} can therefore be evaluated using the definitions of momenta
\eqref{a2} and \eqref{a5} as,

\begin{subequations}\begin{align}
&p_x = \bigg({3\alpha\over \sqrt{2\Lambda}} + {72\Lambda^{3\over 2} f_0\over \sqrt 2}\bigg)\sqrt x
+{V_1\over 4\Lambda^3 a_0\phi_0}x -{\sqrt{\Lambda} a_0^2\phi_0^2\over 2\sqrt 2}{1\over \sqrt x}.\\&
p_z = -\bigg(3\alpha\Lambda + 72\Lambda^3 f_0\bigg)\sqrt z -{V_1 z\over \Lambda a_0\phi_0}
- {\Lambda a_0^2\phi_0^2\over 4}{1\over \sqrt z}.\\&
p_\phi=-\frac{\Lambda a_0^3\phi_0^3}{\phi^2}.\end{align}\end{subequations}
Therefore the form of $S_0(x, z, \phi)$ reads,

\be\label{a10} S_0 = 2 \alpha\Lambda z^{3\over 2} + 48 \Lambda^3 f_0 z^{3\over 2} - {\Lambda\over 2} a_0^2\phi_0^2 \sqrt z.
\ee
Now to find the zeroth order on-shell action, we use classical solution \eqref{sol} to express all the variables in
the action \eqref{A2} or \eqref{a6} as well, in terms of $z$ using \eqref{rel}, and then integrate to obtain
($N = 1,~ k = 0$),

\be\label{A0}\begin{split}A_{2\mathrm{Cl}}& = \int \bigg[ 3\alpha \sqrt z \dot x + {9f\dot x^2 \over \sqrt z}
+ z^{3\over 2}\Big({1\over 2}\dot \phi^2 - V\Big)\bigg]dt \\&
= \int\left[12 \alpha \Lambda^2  +  \left(144\Lambda^4 f_0 + {V_1\sqrt z\over a_0\phi_0}
- {\Lambda^2 a_0^2\phi_0^2\over 2 z}\right) + {\Lambda^2a_0^2\phi_0^2\over 2 z} - \left(6\alpha\Lambda^2
+ {V_1\sqrt z\over a_0\phi_0} + {\Lambda^2 a_0^2\phi_0^2\over 2z}\right)\right]{\sqrt z d z\over 2\Lambda }\\&
= 2\alpha\Lambda z^{3\over 2}+48 \Lambda^3 f_0 z^{3\over 2} - {\Lambda \over 2}a_0^2\phi_0^2\sqrt z.\end{split}\ee
Since Hamilton-Jacobi function \eqref{a10} matches with the zeroth-order on-shell action \eqref{A0}, so everything is consistent. Therefore,
proceeding as before, upto first order of approximation, the semiclassical wavefunction is obtained as

\be \label{a11}\Psi_{2} = \Psi_{02} e^{{i\over \hbar}\Lambda \big[\big(2\alpha+ 48 \Lambda^2 f_0\big) z^{3\over 2}
- {1\over 2}a_0^2\phi_0^2\sqrt z\big]}, ~~~\mathrm{where}, ~~~\Psi_{02} = \psi_{02}e^{H_1(z)}.\ee
Thus, we obtain a wavefunction which is oscillatory about classical inflationary solutions, and therefore is well behaved. So far, it therefore appears that the canonical formulation following HF is mathematically consistent.

\subsection{Modified Horowitz' Formalism (MHF)}

MHF also smoothly bypasses constraint analysis. As repeatedly mentioned, in MHF, we first integrate the action by parts and so in the present case, start with action \eqref{A3} instead. Now introducing an auxiliary variable, $Q = {\partial A_{22}\over \partial \ddot z} = {18 f\over \sqrt z}\Big({\ddot z\over N^3} - {\dot N\dot z\over N^4}\Big)$,
in the action \eqref{A3} and integrating by parts, rest of the total derivative terms viz, $\Sigma_{R_2^2}$ is eliminated, and one obtains

\bes\label{A4} A_{22} = \int\Bigg[ -3\alpha\left({\dot z^2\over 2N\sqrt z} - 2 k\sqrt z\right)-\dot Q\dot z
- {Q^2\sqrt z N^3\over 36 f} -{\dot N\dot z Q\over N}+ {18kf\over \sqrt z}\left({\dot z^2\over N z}
-2{f'\dot\phi\dot z\over Nf}+ 2k N\right) \\& \hspace{3.5 in}+\left({\dot\phi^2\over 2N} -N V\right)z^{3\over 2}\Bigg]dt.
\end{split}\ee
Canonical momenta are

\be \label{cm} P_z = -{3\alpha\dot z\over N\sqrt z} - \dot Q - {\dot N Q\over N} +
{36 k f\over \sqrt z}\Big({\dot z\over N z} -{f'\dot\phi\over Nf}\Big),~P_Q = -\dot z;~P_\phi =
{z^{3\over 2} \dot \phi\over N}- {36k\dot z f'\over N\sqrt z},~P_N = -{\dot z Q\over N}.\ee
The $N$ variation equation is

\be\label{N} -\dot z\dot Q - {3\alpha \dot z^2\over 2N\sqrt z} - {\dot N\dot z Q\over N} +
{18 k f\dot z^2\over N z^{3\over 2}} -{36 k f'\dot \phi\dot z\over N\sqrt z} + {z^{3\over 2}\dot\phi^2\over 2N}
+ N z^{3\over 2}V - 6\alpha kN\sqrt z + {N^2 Q^2\sqrt z\over 36f} - {36k^2 Nf\over \sqrt z}.\ee
Since, $P_Q P_z = \dot z\dot Q + {3\alpha \dot z^2\over N\sqrt z} + {\dot N\dot z Q\over N}
- {36 k f\dot z^2\over N z^{3\over 2}} + {36 k f'\dot \phi\dot z\over N\sqrt z} $, therefore, it is straight forward
to cast the phase-space structure of the Hamiltonian as,

\bes H_{22} = -P_Q P_z +\left({3\alpha\over 2 N\sqrt z} - {18kf\over Nz^{3\over 2}} + {648 k^2 f'^2\over N z^{5\over 2}}\right)
P_Q^2 - {36 k f'\over z^2}P_Q P_\phi +{N\over 2 z^{3\over 2}}P_{\phi}^2 \\& \hspace{2.0 in}+ {N^2 \sqrt z\over 36 f} Q^2
+ Nz^{3\over 2}V(\phi) - 6\alpha k N\sqrt z - {36 k^2 N f\over \sqrt z}.\end{split}\ee
Now to establish diffeomorphic invariance, we express the Hamiltonian in terms of the basic variables. This is performed
by replacing the auxiliary variable $\{Q,~P_Q\}$ to basic variable $\{K_{ij}, ~\Pi^{ij}\}$. For this purpose, we choose
$x = {\dot z\over N}$, so that we need to replace $Q$ by ${P_x\over N}$ and $P_Q$ by $- N x$ to find\footnote{The transformation is canonical, since $N$ should not be treated as a variable, which is transparent under the choice $q = NQ$, as depicted in B1.},

\be\label{H5}\begin{split}& H_{22} = N \mathcal{H}_{22} = \\&N \left[x P_z + {\sqrt z P_x^2\over 36 f}  +
{P_\phi^2\over 2 z^{3\over 2}} + {36 k x f' P_\phi\over z^2} + \left({3\alpha\over 2 \sqrt z} -
{18k f \over z^{3\over 2}} +{648 k^2 f'^2 \over z^{5\over 2}} \right)x^2 - 6\alpha k \sqrt z -
{36k^2 f\over \sqrt z} + V(\phi) z^{3\over 2}\right]. \end{split}\ee
The action \eqref{A3} may be also be expressed in canonical form as before,

\be\label{Can1} A_{22} = \int (\dot x P_x + \dot z P_z + \dot\phi P_\phi - N \mathcal{H}_{22})dt d^3x =
\int \dot h_{ij}P^{ij} + \dot K_{ij}\Pi^{ij} + \dot \phi P_\phi - N \mathcal{H}_{2M})dt d^3x,\ee
where, $P^{ij}$ and $\Pi^{ij}$ stand for the momenta canonically conjugate to $h_{ij}$ and $K_{ij}$ respectively.
Although the two Hamiltonians \eqref{a4} and \eqref{H5} produce the same and unique classical field equations, they
differ from each other by and large. For example, \eqref{a4} contains a linear term in $p_x$, which is absent from
\eqref{H5}. On the contrary, \eqref{H5} contains a linear term in $P_\phi$. Further the effective potentials are
also different. Although, the two are related under the set of transformation relations,

\be z \rightarrow z,~p_z \rightarrow P_z-18 f\frac{k x }{z^{\frac{3}{2}}}+\frac{3\alpha x}{2\sqrt{z}}; ~~x
\rightarrow x,~p_x \rightarrow P_x+36 f \frac{k}{\sqrt{z}}+3 \alpha\sqrt{z};~~\phi \rightarrow \phi,~p_{\phi}
\rightarrow {P_{\phi}+{36 k f' {x\over \sqrt z}}},\ee
which are canonical. However, the difference is revealed from their quantum counterpart. \\

\noindent
\textbf{Canonical quantization:}\\

\noindent
The analogous Schr\"odinger equation corresponding to \eqref{H5} is

\be \label{Sc}\begin{split}\frac{i\hbar}{\sqrt z}\frac{\partial \Psi}{\partial z} =& -\frac{\hbar^2}{36 f x}
\left(\frac{\partial^2 \Psi}{\partial x^2} + \frac{n}{x}\frac{\partial \Psi}{\partial x}\right)-\frac{\hbar^2}{2xz^2}
\frac{\partial^2 \Psi}{\partial\phi^2} + {36 k\over z^{5\over 2}}\widehat{f' P_\phi}\Psi \\&+
\left[\left({3\alpha\over 2 z} - {18k f \over z^2} +{648 k^2 f'^2 \over z^3} \right)x - {36k^2 f\over x z}
+ {z\over x}\big(V(\phi)- 6\alpha k\big)\right]\Psi = 0.\end{split}\ee
Note that while in \eqref{a4}, a linear term in $p_x$ appears and operator ordering is required between $\hat x$ and
$\hat {p}_x$ as done in \eqref{a7}, here in \eqref{H5}, the same appears with $P_\phi$, and one needs to resolve the
operator ordering ambiguity between $\hat f'(\phi)$ and $\hat {P_\phi}$ in \eqref{Sc}. This is possible only after
having knowledge of a specific form of $f(\phi)$. Further the effective potentials are also different by and large.
Therefore the modified Wheeler-deWitt equations are distinct. Now under a further change of variable, and using the form of $f(\phi)$ obtained in \eqref{sol}, the above modified Wheeler-de-Witt equation \eqref{Sc}, takes the look of Schr\"odinger equation, viz.,

\be\begin{split}\label{MHS} i\hbar \frac{\partial \Psi}{\partial\sigma}=&-\frac{\hbar^2}{54f(\phi)}
\left(\frac{1}{x}\frac{\partial^2}{\partial x^2} + \frac{n}{x^2}\frac{\partial}{\partial x}\right)\Psi
-\frac{\hbar^2}{3x\sigma^{\frac{4}{3}}}\frac{\partial^2\Psi}{\partial\phi^2}\\&
+ i\hbar{k\over 4\Lambda^4 \sigma^{5\over 3}}\left[\left({V_1\over \phi^2}
+ \Lambda^2\phi\right){\partial\Psi\over\partial\phi} + \left({\Lambda^2\phi^3
- 2V_1\over 2\phi^3}\right)\Psi\right] +V_e\Psi=\hat{H_e}\Psi\\&
\mathrm{where},~~V_e = \left({3\alpha\over 2 \sigma^{2\over 3}} - {18k f \over \sigma{^4\over 3}}
+{648 k^2 f'^2 \over \sigma^2} \right)x  - {36k^2 f\over x \sigma^{2\over 3}} + {\sigma^{2\over 3}\over x}\big(V(\phi)
-6\alpha k\big).
\end{split}\ee
In the above, the proper volume, $\sigma=z^{\frac{3}{2}}=a^3$ plays the role of internal time parameter, and we have
performed the Weyl symmetric operator ordering between $\hat f'(\phi)$ and $\hat P_\phi$. Further, $\hat H_e$ and $V_e$
are the effective Hamiltonian operator and the effective potential respectively. The hermiticity of the effective
Hamiltonian is ensured for $n = -1$, which enables one to write the continuity equation as,

\be   \frac{\partial\rho}{\partial\sigma}+\nabla.\mathrm{\mathbf{J}}=0,\ee
where, $\rho = \Psi^*\Psi$ is the probability density and $\mathrm{\mathbf{J}} = ({\mathrm{J}_x, \mathrm{J}_\phi},0)$
is the current density, with, $\mathrm{J}_x = {i\hbar\over 54 f x} (\Psi^*_{,x} \Psi - \Psi^*\Psi_{,x})$
and $\mathrm{J}_\phi = {i\hbar\over 3 x \sigma^{4\over 3}}(\Psi^*_{,\phi} \Psi - \Psi^*\Psi_{,\phi})
- {k\over 4\Lambda^4 \sigma^{5\over 3}}\left[\left({V_1\over \phi^2} + \Lambda^2\phi\right)\Psi^*\Psi\right]$.
In the process, operator ordering ambiguity is resolved ($n = -1$) from physical consideration.\\

\noindent
\textbf{Semiclassical approximation:}\\

\noindent
As before, in order to perform semiclassical approximation, we express the modified Wheeler-deWitt equation ($k = 0, n = -1$) as,

\be\label{Sck}\frac{i\hbar}{\sqrt z}\frac{\partial \Psi}{\partial z} = -\frac{\hbar^2}{36 f x}
\left(\frac{\partial^2}{\partial x^2} + \frac{n}{x}\frac{\partial}{\partial x}\right)\Psi-
\frac{\hbar^2}{2xz^2}\frac{\partial^2\Psi}{\partial\phi^2}  + \left[{3\alpha x\over 2 z}
+  {zV\over x} \right]\Psi = 0.\ee
which may be further rearranged to obtain,

\bes \label{WE2}-\frac{\hbar^2\sqrt z}{36 x}\left(\frac{\partial^2}{\partial x^2}
+ \frac{n}{x}\frac{\partial}{\partial x}\right)\Psi
-f(\phi)\left[\frac{\hbar^2}{2xz^{3\over 2}}\frac{\partial^2\Psi}{\partial\phi^2}
+i\hbar\frac{\partial \Psi}{\partial z}\right] + f(\phi)\left[{3 \alpha x\over 2\sqrt z}
+ {z^{3\over 2}V\over x}\right]\Psi = 0.\end{split}\ee
The above equation may be viewed as time independent Schr\"{o}dinger equation with three variables $x$, $z$ and $\phi$.
Hence, let us again seek the solution of the wavefunction as, and insert \eqref{Psi1} and \eqref{psi2} in the wave-equation \eqref{WE2} one obtains

\bes
\Bigg[-{\hbar^2 \sqrt z\over 36 x}\Bigg\{{i\over \hbar}\big[S_{0,xx} + \hbar S_{1,xx} + \hbar^2 S_{2,xx}
+ \mathcal{O}(\hbar)\big]\\& -{1\over \hbar ^2}\big[S_{0,x}^2 + \hbar^2 S_{1,x}^2  + 2\hbar S_{0,x} S_{1,x}
+ 2\hbar^2 S_{0,x} S_{2,x} +\mathcal{O}(\hbar)\big]
+{i n\over \hbar x}\big[S_{0,x}+\hbar S_{1,x}+\hbar^2 S_{2,x}+\mathcal{O}(\hbar)\big]\Bigg\}\\&
-{f(\phi)\hbar^2\over 2x z^{3\over 2}}\left\{{i\over \hbar}\big[S_{0,\phi\phi} + \hbar S_{1,\phi\phi}
+ \hbar^2 S_{2,\phi\phi}+\mathcal{O}(\hbar)\big] -{1\over \hbar ^2}\big[S_{0,\phi}^2 + \hbar^2 S_{1,\phi}^2
+ 2\hbar S_{0,\phi} S_{1,\phi} + 2\hbar^2 S_{0,\phi} S_{2,\phi} +\mathcal{O}(\hbar)\big]\right\}\\&
- i\hbar f(\phi) \left\{{i\over \hbar}\big[S_{0,z}+\hbar S_{1,z}+\hbar^2 S_{2,z}+\mathcal{O}(\hbar)\big]\right\}
+ f(\phi)\left\{{3 \alpha x\over 2\sqrt z} + {z^{3\over 2}V\over x}\right\}\Bigg]\Psi=0\end{split} .\ee
Finally, equating the coefficients of different powers of $\hbar$ to zero, the following set of equations
(upto second order) is obtained.

\begin{subequations}\begin{align}
&\label{001}\frac{\sqrt z}{36 x}S_{0,x}^2+f(\phi)\frac{S_{0,\phi}^2}{2xz^{\frac{3}{2}}}+ f(\phi)S_{0,z}
+ f(\phi)\left({3 \alpha x\over 2\sqrt z} + {z^{3\over 2}V\over x}\right)= 0\\
&\label{002}-\frac{i\sqrt z}{36 f x}S_{0,xx}-\frac{in\sqrt z}{36 f x^2}S_{0,x}-\frac{iS_{0,\phi\phi}}{2xz^{\frac{3}{2}}}
+S_{1,z}+\frac{\sqrt zS_{0,x}S_{1,x}}{18 f x}+\frac{S_{0,\phi}S_{1,\phi}}{xz^{\frac{3}{2}}}=0\\
&-\frac{i\sqrt z}{36 f x}S_{1,xx}-\frac{in\sqrt z}{36 f x^2}S_{1,x}-\frac{iS_{1,\phi\phi}}{2xz^{\frac{3}{2}}}+S_{2,z}
+\frac{\sqrt zS_{0,x}S_{2,x}}{18 f x}+\frac{S_{0,\phi}S_{2,\phi}}{xz^{\frac{3}{2}}} + \frac{\sqrt z}{36 f x}S_{1,x}^2
+\frac{S_{1,\phi}^2}{2x z^{3\over 2}}=0
\end{align}\end{subequations}
which are to be solved successively to find $S_0(x, z,\phi)$, $S_1(x, z,\phi)$ and $S_2(x, z,\phi)$ and so on. Now
identifying $S_{0,x}$ with $P_x$, $S_{0,z}$ with $P_z$ and $S_{0,\phi}$ with $P_\phi$, one can obtain the Hamilton
constraint equation \eqref{H5}. Further in view of the definition of canonical momenta $P_x$ and $P_{\phi}$ \eqref{cm}
and $P_z = -\dot P_x$, it is also possible to regain the time-time component of Einstein's equation \eqref{FE}
in flat space. So everything so far is consistent. Now, in order to compute $S_0(x, z, \phi)$, let us express it as,

\be\label{S3} S_0=\int P_z dz+\int P_x dx+\int P_\phi d\phi\ee
apart from a constant of integration which may be absorbed in $\Psi_0$. In view of the classical solution \eqref{sol},
it is possible to interrelate all the variables. For example,

\bes\label{Relate} x = \dot z = 2\Lambda z;~\ddot z = 4\Lambda^2 z;~\dddot z = 8\Lambda^2 z;~~ \dot\phi =
-\Lambda \phi;~ \phi = {a_0\phi_0\over \sqrt z},\\& f = f_0 +{V_1\over 144\Lambda^4}{1\over \phi} -
{\phi^2\over 288\Lambda^2} = f_0 + {V_1\over 144\Lambda^4 a_0\phi_0}\sqrt z - {a_0^2\phi_0^2\over 288 \Lambda^2 z}
= f_0 + {V_1\over 144\Lambda^4 a_0\phi_0}\sqrt{x\over 2\Lambda} - {a_0^2\phi_0^2\over 144 \Lambda x}.\end{split}\ee
Canonical momenta \eqref{cm} therefore take the following forms,

\begin{subequations}\begin{align}\label{MOM}& P_x = Q = 18 f{\ddot z\over\sqrt z} =
{72\Lambda^2\sqrt x\over \sqrt 2\Lambda} \bigg(f_0 + {V_1 \sqrt x\over 144\Lambda^4 a_0\phi_0\sqrt {2\Lambda}}
- {a_0^2\phi_0^2\over 144 \Lambda x}\bigg) ={72\Lambda^{3\over 2}f_0\sqrt x\over \sqrt 2}
+ {V_1 x\over 4 \Lambda^3 a_0\phi_0} - {{\sqrt\Lambda} a_0^2\phi_0^2\over 2 \sqrt {2x}}\\&
P_z = - 6\alpha\Lambda\sqrt z - 72\Lambda^3 f_0\sqrt z - {V_1 z\over \Lambda a_0\phi_0}
- {\Lambda a_0^2\phi_0^2\over 4\sqrt z} \\&
P_\phi = z^{3\over 2}\dot \phi = -\Lambda {a_0^3\phi_0^3\over \phi^2}
\end{align}\end{subequations}
Hence

\be \label{S0M} S_0 = \int(P_x dx + P_z dz + P_\phi d\phi) = - 4\alpha\Lambda z^{3\over 2} +48 \Lambda^3 f_0 z^{3\over 2}
- {\Lambda\over 2} a_0^2\phi_0^2\sqrt z .\ee
One can also compute the form of the zeroth order on-shell action ($k = 0, N = 1$) in view of the canonical action \eqref{A3} or
\eqref{Can1} and the Hamiltonian \eqref{H5} as well, using the interrelations between the variables \eqref{Relate} and the
definition of momenta \eqref{MOM} as

\bes\label{CLAS} A_{22\mathrm{Cl}} = \int\left[\dot x P_x + \dot\phi P_\phi - {\sqrt z\over 36 f}P_x^2 -
{P_\phi^2\over 2 z^{3\over 2}} - \frac{3\alpha x^2}{2\sqrt z} - V(\phi) z^{3\over 2}\right]dt\\& =
\int\left[- 12 \alpha \Lambda^2 z^{3\over 2} + 144\Lambda^4 z^{3\over 2}\left(f_0 + {V_1\over 144\Lambda^4\phi}
-{\phi^2\over 288\Lambda^2}\right) - {V_1 z^2\over a_0\phi_0}\right]{dz\over 2\Lambda z}\\& \int
\left[- 12 \alpha \Lambda^2 z^{3\over 2} + 144\Lambda^4 f_0z^{3\over 2} -{\Lambda a_0^2\phi_0^2\over 2}\sqrt z\right]
{dz\over 2\Lambda z} = - 4\alpha\Lambda z^{3\over 2} +48 \Lambda^3 f_0 z^{3\over 2}
- {\Lambda\over 2} a_0^2\phi_0^2\sqrt z .\end{split}\ee
Hence, the above zeroth order on shell action \eqref{CLAS} matches here again with Hamilton-Jacobi function \eqref{S0M}. Nevertheless, one may note the difference between the Hamilton-Jacobi functions obtained following HF \eqref{a10} and MHF \eqref{S0M}, which distinguish the two Hamiltonian as distinct in the quantum domain. As before, it is also possible to find the semiclassical wavefunction upto first order approximation and the result is,

\be \label{ScWd}\Psi_{2M} = \Psi_{022} e^{{i\over \hbar}\Lambda \big[\big(-4\alpha+ 48 \Lambda^2 f_0\big) z^{3\over 2}
- {1\over 2}a_0^2\phi_0^2\sqrt z\big]}, ~~~\mathrm{where}, ~~~\Psi_{022} = \psi_{022}e^{H_2(z)}.\ee
The wavefunction here again executes oscillatory behaviour about classical inflationary solutions, and therefore is
well behaved.\\

\end{document}